\documentclass[pra,twocolumn, showpacs, superscriptaddress,preprintnumbers,amsmath,amssymb,longbibliography]{revtex4-2}  

\AtBeginDocument{\usepackage{booktabs}}

\usepackage{amsmath}
\usepackage{amssymb}
\usepackage{physics}
\usepackage{color}
\usepackage{comment}

\usepackage{multirow}
\usepackage{makecell}

\usepackage[hidelinks]{hyperref}

\usepackage{graphicx}  
\usepackage{dcolumn}   
\usepackage{bm}        
\usepackage{amssymb}   
\usepackage{natbib}
\usepackage{bm}
\usepackage{tabularx}

\usepackage{ulem}
\usepackage{hyperref}

\newcommand{\beq}{\begin{equation}}
\newcommand{\eeq}{\end{equation}}

\DeclareUnicodeCharacter{2032}{$^\prime$}
\DeclareUnicodeCharacter{0085}{\\}
\begin{document}

\title{Dynamical control in a prethermalized molecular ultracold plasma: Local dissipation drives global relaxation }

\author{Ruoxi Wang}
\affiliation{Department of Physics \& Astronomy, University of British Columbia, Vancouver, BC V6T 1Z3, Canada}
\author{Amin Allahverdian}
\affiliation{Department of Chemistry, University of British Columbia, Vancouver, BC V6T 1Z1, Canada}
\author{Smilla Colombini}
\affiliation{Department of Chemistry, University of British Columbia, Vancouver, BC V6T 1Z1, Canada}
\author{Nathan Durand-Brousseau}
\affiliation{Department of Physics \& Astronomy, University of British Columbia, Vancouver, BC V6T 1Z3, Canada}
\affiliation{Department of Chemistry, University of British Columbia, Vancouver, BC V6T 1Z1, Canada}
\author{Kevin Marroqu\'in }
\affiliation{Department of Chemistry, University of British Columbia, Vancouver, BC V6T 1Z1, Canada}
\author{James Keller}
\affiliation{Department of Chemistry, Kenyon College, Gambier, OH 43022 USA}
\author{John Sous}  
\email[Electronic mail:  ]
{jsous@ucsd.edu}
\affiliation{Department of Chemistry, University of California, San Diego, La Jolla, CA 92093 USA }
\author{Abhinav Prem}
\email[Electronic mail:  ]
{aprem@ias.edu}
\affiliation{School of Natural Sciences, Institute for Advanced Study, Princeton, New Jersey 08540, USA}
\author{Edward Grant}
\email[Electronic mail:  ]
{edgrant@chem.ubc.ca}
\affiliation{Department of Physics \& Astronomy, University of British Columbia, Vancouver, BC V6T 1Z3, Canada}
\affiliation{Department of Chemistry, University of British Columbia, Vancouver, BC V6T 1Z1, Canada}

\begin{abstract}

Prethermalization occurs as an important phase in the dynamics of isolated many-body systems when strong coupling drives a quasi-equilibrium in a subspace separated from the thermodynamic equilibrium by the restriction of a gap in energy or other conserved quantity.  Constrained relaxation in a prethermal phase can support restricted mobility in natural and model systems of high dimensionality and limited disorder.  Here, we report the signature of an enduring prethermal regime of arrested relaxation in the molecular ultracold plasma that forms following the avalanche of a state-selected Rydberg gas of nitric oxide.  For a wide range of initial conditions, this system enters a critical phase in which a density of NO$^+$ and electrons balances a population of Rydberg molecules.  Electron collisions mix orbital angular momentum, scattering Rydberg molecules to states of very high-$\ell$.  Spontaneous predissociation purifies this non-penetrating character, creating an extraordinary gap between the plasma states of $n \approx \ell$, with measured $n>200$ and penetrating states of $\ell = 0, ~1$ and 2.  Evolution to a statistically equilibrated state of N and O atoms cannot occur without Rydberg electron penetration, and this gap blocks relaxation for a millisecond or more.  Evolving through the critical regime, electrons that balance the NO$^+$ charge behave as though localized in the prethermal phase and play an ineffective role in bridging this gap.  However, the application of a weak radiofrequency (RF) field promotes a dramatic degree of relaxation owing to electron collisions.  On an entirely different scale, exciting a quantum-state transition in an exceedingly small fraction of the molecules in the prethermalized ensemble acts with even greater effect to drive the entire system toward equilibrium.  We ascribe this to an introduction of dissipative character to a small fraction of the states in the prethermally localized ensemble.  In time, this character spreads to bridge the angular momentum gap and cause the entire ensemble to thermalize.  Using the Lindblad master equation, we illustrate qualitatively similar dynamics for a toy model of an open quantum system that consists of a localized set of spins on which dissipation acts locally only at a single site.

\end{abstract}


\maketitle

\section{Introduction}

The collective dynamics of many interacting degrees of freedom can give rise to rich non-equilibrium phenomena, such as classical synchronization and chaos under unitary quantum evolution, with important implications for physical, chemical, and biological systems~\cite{nitsan2016mechanical,witthaut2017classical,eneriz2019degree}.  Ordinarily, one expects interactions in a macroscopic system to drive relaxation to thermal equilibrium because different subsystems act as reservoirs for each other.  However, recent progress in understanding isolated quantum systems evolving under unitary dynamics has led to the discovery of striking examples of non-equilibrium physics that fail to conform with the notion of thermalization according to the eigenstate thermalization hypothesis (ETH)~\cite{deutsch1991,srednicki1994,rigol2008,polkovnikov2011,rigol2016}  

Aside from the novel phenomena of quantum many-body scarring~\cite{shiraishi2017systematic,moudgalya2018a,moudgalya2018b,turner2018quantum,schecter2019weak,spielman2022slow} and Hilbert space fragmentation~\cite{khemani2019local,sala2020ergodicity,moudgalya2019krylov,rakovszky2020statistical}, many-body localization (MBL)~\cite{gornyi2005,baa2006,pal2010} stands out as a canonical example of ergodicity breaking in interacting nonintegrable quantum systems. In contrast to integrable cases, in which fine-tuning yields conserved quantities, MBL arises from a complex interplay of interactions and disorder that gives rise to extensively many emergent local integrals of motion, resulting in remarkable dynamics (see Refs.~\cite{huse2015review,altman2015review,abanin2018review} for a review).

A great deal of theoretical work has explored the possibility of MBL in the presence of long-range interactions~\cite{yao2014many,wu2016,nandkishore2017mbl,mirlin2018,nag2019}, with the aim in part to enable its realization in practical quantum devices. However, even in one spatial dimension, the clearest theoretical evidence for MBL -- as strictly defined in the limit of long times and large volumes -- appears in short-range interacting strongly disordered systems.  Long-range interactions lead invariably to thermalization at long-times~\cite{de2017stability,luitz2017,thiery2018avalanche}. 

Thus, one should not expect a three-dimensional system to form an MBL phase, even if confined to short-range interactions~\cite{morningstar2022avalanches}.  Instead, realistic disorder and ubiquitous long-range interactions are much more likely to produce a condition of partitioned entanglement~\cite{monteiro2021quantum,long2023phenomenology}.   In transient regimes out of equilibrium, such variations in the strength of local interactions cause observables to relax on different timescales.   

This can drive interacting degrees of freedom in an isolated quantum many-body system to a \textit{prethermalized} state of quasi-equilibrium, such that the entire system evolves via a long-lived metastable state to eventually reach the state determined by statistical thermodynamics~\cite{berges2004prethermalization,gring2012relaxation,langen2016prethermalization,mallayya2019prethermalization,WOS:000720068200007}.  Indeed, unlike other forms of ergodicity-breaking, prethermalization is \textit{generically} expected in a large class of systems in \textit{any} spatial dimension~\cite{yin2023prethermalization}, making it an important dynamical regime to study both experimentally and theoretically.

More generally, prethermalization stemming from constrained quantum dynamics has deep connections with various emergent phenomena, including glassiness~\cite{markland2011quantum,olmos2014out,everest2016emergent}, quantum magnetism~\cite{glaetzle2014quantum,zeiher2016many,harris2018phase,koepsell2019imaging}, thermalization in anyonic spin chains~\cite{chandran2016eigenstate,chen2018does}, disorder-free localization~\cite{deroeck2014trans,deroeck2014AMBL,groverfisher,schiulaz2015,papic2015nodisorder,yao2016,chamon2005,prem2017glass,smith2017a,smith2017b,michailidis2018,brenes2018,sous2021phonon}, and out-of-equilibrium superconductivity~\cite{mitrano2016possible,giannetti2016ultrafast,basov2017towards,tokura2017emergent,kennes2017transient}. 

The characteristics necessary for prethermalization typically include the constraint of an energy gap or some other conserved quantity that regulates the progress of an observable to its equilibrated distribution~\cite{mori2018thermalization,yin2023prethermalization,mori2024liouvillian}. Isolated systems that evolve to a prethermal stage often exhibit a non-analytic order parameter, signaling the critical point associated with a dynamical phase transition~\cite{WOS:000870855000001}. 

However, the complexity of interacting many-body systems, which evolve in Hilbert spaces that depend exponentially on the system size, renders their prethermal dynamics inaccessible to existing analytical or numerical techniques, particularly in three dimensions.  This has motivated concomitant experimental efforts to fashion small ensembles as quantum simulators and annealers~\cite{hauke2020perspectives,winci2020path}, with well-isolated many-body networks of ultracold atoms--engineered to arrange positions and tune interaction energies--having successfully demonstrated striking dynamical outcomes~\cite{gross2017quantum}. While such quantum simulators offer access to single-site resolution as well as system sizes on the order of tens of particles, which are beyond those addressable by large-scale numerics, finite coherence times limit the time scales over which they can represent out-of-equilibrium dynamics.  

Rydberg gases offer particular advantages in this arena as materials from which to form out-of-equilibrium strongly correlated many-body systems. Rydberg-Rydberg interactions occur with exaggerated coupling energies in systems of varied density \cite{browaeys2016experimental}. Cooperative behaviour in atomic Rydberg ensembles ranges from precisely defined pairwise and higher-order coherent phenomenon \cite{vsibalic2016driven,firstenberg2016nonlinear,gambetta2020engineering,browaeys2020many}  to aggregation \cite{garttner2013dynamic,Schempp2014,urvoy2015strongly}, dissipation \cite{lesanovsky2014out,goldschmidt2016anomalous,letscher2017bistability,bernien2017probing}, nonequilibrium phase transitions \cite{carr2013nonequilibrium,malossi2014full} and avalanche to plasma \cite{Killian2007,weller2019interplay,review}. Recent experiments have established evidence for self-organizing dynamics in optically driven ultracold and room-temperature atomic Rydberg gases \cite{helmrich2020signatures,ding2020phase} and in the spontaneous evolution of a molecular ultracold plasma \cite{marroquin2024self}.

In this work, we explore non-equilibrium dynamics in a long-range interacting, disordered many-body Rydberg gas of nitric oxide molecules and find remarkable evidence for a localized prethermal state which, absent any external perturbations, remains stationary over observation time-scales. Specifically, we investigate the dynamics of the out-of-equilibrium state formed when a Rydberg gas of nitric oxide entrained at a milli-Kelvin temperature in a skimmed supersonic molecular beam undergoes an electron-impact avalanche to form a strongly coupled molecular ultracold plasma \cite{Plasma_prl,Sadeghi:2014,schulz2016evolution,Schulz-Weiling2016,Haenel2017,Haenel.2018}.  As this system evolves, hydrodynamic forces combine with dissipation in a molecular NO$^+$ ion - electron, NO$^*$ Rydberg excited-state landscape, driving self-organization to sustain a long-lived prethermalized state.  

This state settles for long observation times in a condition of universal density with an exceedingly small electron-ion binding energy and high orbital angular momentum, $\ell$.  Strikingly, these prethermal characteristics depend very little on the initial density or quantum state-selected initial energy of the Rydberg gas \cite{marroquin2024self}. Instead, we find compelling evidence that the prethermal state results from an anomalously large gap in electron orbital angular momentum, $\ell$, between high-$\ell$ plasma states ($n \approx \ell > 200$) and penetrating states (with $\ell = 0,1,2$).  

Prior work has focused on the universal long-time behavior of this plasma and aspects of localization relating to its intrinsic disorder \cite{Sous2018,Sous2019}.  Here, we show that predissociation plays an essential role in its stability by creating the emergent angular momentum gap that gives rise to prethermalization in this large many-body long-range interacting molecular system. 

In the absence of external perturbations, this prethermalized state survives unchanged for many hundreds of microseconds despite readily available channels of decay to ground-state neutral atoms \cite{Schulz-Weiling2016,Haenel2017}.  This durability provides strong evidence for a prethermal state that is \textit{localized} on an observational timescale, and characterized by electronic degrees of freedom whose mobility is strongly constrained by intrinsic disorder despite the presence of long-range interactions.  Nevertheless, the introduction of any of several weak forms of perturbation that function to bridge the gap between non-penetrating states of high-$\ell$ and predissociative low-$\ell$ states in a small fraction of the high-Rydberg population propagates a large-scale population loss \cite{wang2020radio,wang2022mm}.  

We build upon a conceptual model for arrested relaxation in an ultracold plasma state of randomly interacting dipoles of random energies \cite{Sous2018,Sous2019} to describe field-induced and electromagnetic $\ell$-mixing as a vehicle for adding dissipative states to the prethermalized ensemble.  The predissociative character of those states couples the ensemble to the fully thermal continuum of ${\rm{N (^4S) + O (^3P)}}$ atoms.  We model these dynamics by a Lindblad master equation that describes a system of localized spins dissipatively coupled to a Markovian reservoir at only a single site, which qualitatively reproduces our observations.  These findings support the notion of the ultracold molecular plasma as a novel experimental platform on which to base the study of disordered dynamics and relaxation in a dissipative many-body system.

\section{Experimental methods}

\subsection{Preparation of Rydberg gas}

Nitric oxide, seeded 1:10 in a supersonic expansion of He, cools to moving-frame longitudinal and cross-beam translational temperatures of $T_{\parallel} = 0.5$ K and $T_{\perp} = 5$ mK \cite{schulz2016evolution}.  This jet expands from a stagnation pressure of 500 kPa, entering the source chamber with a background pressure of $10^{-7}$ Torr.  Propagating in $z$, it passes through a 1 mm diameter skimmer to form a molecular beam in the experimental chamber, which is held to a pressure below $10^{-8}$ Torr. 

A pair of co-propagating frequency-doubled 5 ns Nd:YAG pumped dye laser pulses ($\omega_1 + \omega_2$) cross the  NO molecular beam between two grids G$_1$ and G$_2$ in the experimental chamber.  Double-resonant excitation forms a state-selected Rydberg gas ellipsoid with Gaussian dimensions, $\sigma_{yz} = 0.5$ mm and $\sigma_{x} = 1.5$ mm, determined respectively by the diameter of the leading laser pulse, $\omega_1$, and the width of the molecular beam intersected 35 mm after the skimmer.  In the center of this Gaussian ellipsoid, a high-density ultracold Rydberg gas evolves in the 1400 ${\rm ms^{-1}}$ moving frame of the molecular beam.  

\begin{figure}[h!]
\centering
\includegraphics[scale=0.50]{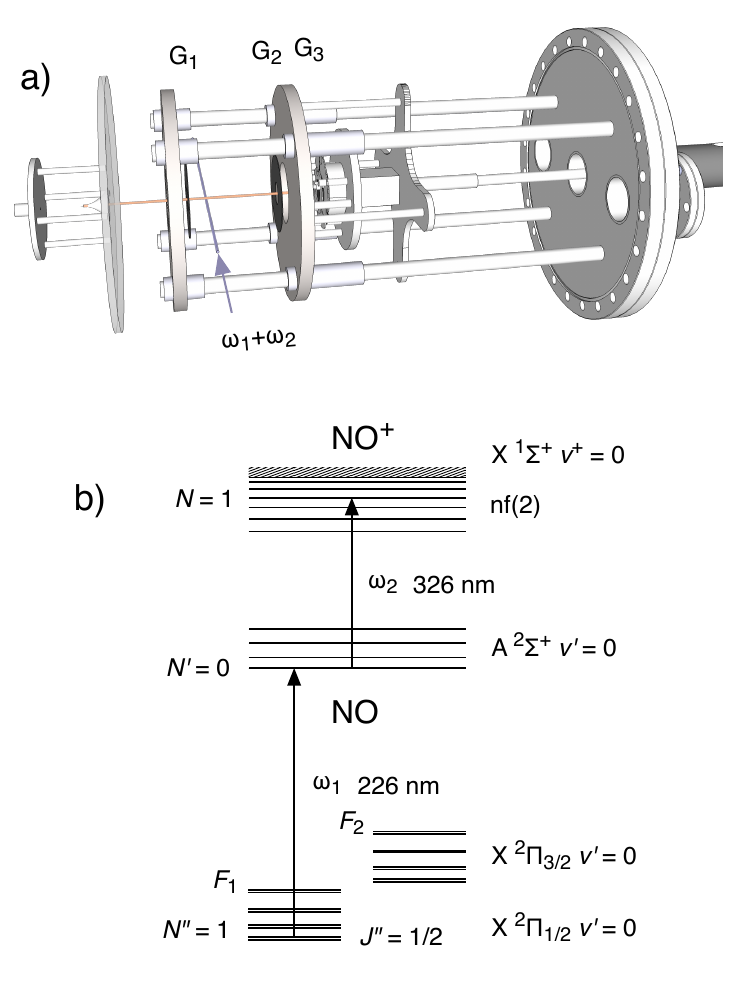}
\caption{\small {(a) Co-propagating laser beams, $\omega_1$ and $\omega_2$, cross a molecular beam of nitric oxide between entrance aperture G$_1$ and grid G$_2$ of a differentially-pumped vacuum chamber. G$_2$, G$_3$ and a MCP detector assembly move on a carriage to afford externally adjustable G$_1$ - G$_2$ spacings from 3 to 60 mm (b)  Nd:YAG-pumped dye lasers ($\omega_1$ and $\omega_2$)  pump ground state NO first to the excited A$ ^2 \Sigma^+ \ N^{'} =0$ state and then to a Rydberg level with $N=1$. }}
\label{fig:diagram}
\end{figure}

The $\omega_1$ pulse excites ground state NO molecules to rotationally selected electronically excited intermediate states, the spectrum of which measures a rotational temperature as low as 2.5 K. For present purposes, $\omega_1$ fixed on a wavelength near 226 nm excites X $^2\Pi_{1/2}$ $N^{''}=1$ $\to$ A $^2\Sigma^{+} $ $N^{'}=0$. The pulse energy of $\omega_1$ is held between 1 and 6 $\mu$J, a point at which the A $\leftarrow$ X transition saturates.   

The $\omega_2$ pulse fully overlaps the population of A $^2\Sigma^{+}$ $N^{'}=0$ molecules photoselected by $\omega_1$.   Scanning the wavelength of $\omega_2$ from 329.0 to 327.6 nm drives subsequent transitions to a series of high-Rydberg states with principal quantum numbers extending from $n_0 = 30$ to 80 and total angular momentum neglecting spin strictly confined to $N=1$.  An output pulse energy of 6 $\mu$J suffices to saturate this $\omega_2$ transition for any selected principal quantum number in this range. 

Thus, for a fixed $\omega_1$ pulse energy, the fraction of A $^2\Sigma^{+} $ $N^{'}=0$ intermediate states that remain after any $\omega_1-\omega_2$ delay precisely determines the relative density of Rydberg states.  Under typical experimental conditions, laser pulses intersecting the molecular beam 35 mm from the nozzle with sufficient intensity to saturate $\omega_1$ and $\omega_2$ transitions form a Rydberg gas ellipsoid with a peak density of $5 \times 10^{12}$  cm $^{-3}$.  

Fixing the $\omega_1$ energy, we vary the $\omega_1-\omega_2$ delay to tune the Rydberg gas density by factor $e^{-\Delta t/\tau}$, where $\tau=196$ ns is the radiative lifetime of the $^2\Sigma^{+} $ $N^{'}=0$ state of nitric oxide \cite{few2014rate,settersten2009collisional}.  Increasing the $\omega_1-\omega_2$ delay from 0 to 1 $\mu$s reduces the initial density of Rydberg gas from $5 \times 10^{12}$ cm$^{-3}$ to $1 \times 10^{10}$ cm$^{-3}$.

\subsection{Low-field propagation and the long-time measure of plasma evolution }

Following $\omega_1-\omega_2$ photo preparation, the Rydberg gas, evolving to plasma, propagates with the molecular beam from the point of illumination near G$_1$ until it reaches G$_2$ where it crosses a grid that shields a field-ionization and electron collection region between G$_2$ and G$_3$.  A bellows actuator longitudinally translates a carriage holding G$_2$, G$_3$ and a microchannel plate detector assembly (Figure~\ref{fig:diagram}a).  Adjusting the flight distance between the $\omega_1-\omega_2$ interaction region and G$_2$ from 3 to 60 mm varies the flight time from 2 to 40 $\mu$s between preparation of the Rydberg gas ellipsoid and the detection of electrons extracted from the ultracold plasma.   

With G$_1$ grounded, a potential applied to G$_2$ varies the longitudinal field in the experimental region between G$_1$ and G$_2$.  The evident polarization of electrons bound by the plasma space charge affords a means to accurately gauge the condition of zero-field (see below).  The experiment deliberately sets confirmed conditions of positive, zero or negative potential in the range from +1 to -1 V/cm, defined as forward, zero and reverse bias with reference to an observed effect on plasma electrons in the illuminated volume.  Experiments that study the waveform effects of forward or reverse bias as a function of plasma flight time adjust the potential applied to G$_2$ to maintain a constant field between G$_1$ and G$_2$.   

As the plasma volume transits G$_2$, it encounters a static field of 200 V cm$^{-1}$ maintained between G$_2$ and G$_3$.  This field extracts an electron signal that tracks the longitudinal width of the illuminated volume at the flight distance specified by the position of G$_2$.  A dual microchannel plate detector proportionally measures the magnitude of this signal, which is recorded as a function of time on an Alligent 4104 digital oscilloscope.  

For the purposes of a few comparative measurements, we used a long flight-path imaging ultracold plasma molecular beam spectrometer \cite{schulz2016evolution,Schulz-Weiling2016,marroquin2024self}.  This apparatus features a field-free flight path of 500 mm from the point at which $\omega_1$ and $\omega_2$ laser pulses cross a skimmed molecular beam to the transverse grids of a selective field-ionization spectrometer between which a long-lived plasma passes before encountering a 75 mm multichannel detector with a phosphor screen imaging anode.

\subsection{Selective field ionization}

An electrostatic field ramp applied between the time of $\omega_2$ ($t_{\omega_2}$ and the appearance of the illuminated volume at G$_2$ forms a selective field ionization (SFI) spectrum that probes the distribution of electron binding energy in the evolving ultracold plasma at the time of the ramp, $t_r$.  We form this ramp by triggering a Bellke high voltage push-pull switch biased by a Bertran high voltage power supply to generate a -3 kV high voltage pulse on G$_1$.  A 5 k$\Omega$ resistor in series with G$_1$ affords an RC circuit that forms a ramped electric field between G$_1$ and G$_2$ with a rise time of $\sim 0.8 $ V cm$^{-1}$ns$^{-1}$. 

We precisely fit a polynomial function to the leading edge of this voltage pulse, transforming the time-dependent electron signal waveform to a quantity that varies with the electric field in V/cm.  Electric field ramps started at chosen particular delays after $\omega_2$ produce SFI spectra of electron binding energies.  We typically collect sets of 4,000 individual SFI spectra at each ramp-field delay.  For a fixed ramp delay after $\omega_2$ excitation, we raise the $\omega_1$ pulse energy to 6 $\mu$J, saturating the first-photon transition.  A systematic variation of the time between $\omega_1$ and $\omega_2$ surveys Rydberg gas Gaussian ellipsoids over a large range of scale as measured by the major and minor radii of shells of specified density.

\subsection{Sources for time-dependent RF and mm-wave excitation spectroscopy }

A Rhode and Schwarz signal generator (SMB100B) synthesizes a radiofrequency pulse in the range from 8 kHz to 6 GHz with arbitrary pulse width and delay and amplitude as high as 5 V peak-to-peak.  Experiments probing the effect of an RF field on plasma dissipation apply this pulse to grid, G$_1$ or G$_2$.  For the purpose of the present measurements, a LabVIEW interface sets the frequency to 60 MHz, with an amplitude of 400 mV cm$^{-1}$ and pulse duration of 250 ns triggered at a time, $\Delta t_{\omega_{\rm RF}}$ after $t_{\omega_2}$.   At the zero-crossing, this field adds $\pm 9.3 \times 10^3$ m s$^{-1}$ to the electron velocity, which corresponds to an energy of 0.25 meV or 60 GHz.  The amplitude of electron oscillation is $\pm 25$ $\mu$m, which is less than 10 percent of the plasma radius. 

This synthesizer also serves as the frequency source for mm-wave electromagnetic spectroscopy.  For this purpose, a Rhode and Schwarz field of selected single-frequency seeds a Virginia Diodes low-frequency sideband generator system (VDI-MixAMC-S116), which produces an output frequency equal to the input frequency plus 10.6 GHz.  This yields a waveform that oscillates at a precisely selected frequency in the range from 11 GHz to 17 GHz.  

This primary frequency passes to a Virginia Diodes 10-400 GHz solid-state multi-band transmitter (VDI-Tx-S129).  Frequency doubling (WR 5.1$\times$2R2) and tripling (WR 9.3$\times$3) stages produce an output mm-wave field, $\omega_{\rm mm}$, that is continuously tunable from 66 to 100 GHz. The timing of $\omega_{\rm mm}$ is precisely controlled via the input synthesizer signal.  The multi-band transmitter broadcasts this output through a VDI-WR-10 35.5$\times$16.3 mm conical horn with an {intensity} as high as 100 $\mu$W cm$^{-2}$.  

An $f/5$ Teflon lens collects a portion of this mm-wave electromagnetic field and focuses it on the molecular beam with a maximum estimated {intensity} of 25 $\mu$W cm$^{-2}$. Given the diameter of the vacuum chamber window through which the mm-wave beam is collimated and the speed of the supersonic beam, the mm-wave field overlaps a given volume of the molecular beam for a maximum of 20 $\mu$s.  

By tuning $\omega_1$ and $\omega_2$ to a particular Rydberg state and scanning the mm-wave frequency, we can detect similar spectra both in the SFI and late peak ultracold plasma signals when the three radiation fields satisfy the triple-resonance condition \cite{wang_2022}. We modulate the synthesized input signal to delay or extend a pulsed application of the mm-wave field $\omega_{\rm mm}$ after the $\omega_2$ excitation that forms the nitric oxide Rydberg gas.

\section{Results}

\subsection{The avalanche of a state-selected Rydberg gas of nitric oxide to form a molecular ultracold plasma }

The $\omega_1$, $\omega_2$, double resonant sequence of laser pulses prepares a Rydberg gas of specified initial density, $\rho_0$, and principal quantum number $n_0$.  The first laser drives a transition to the A $^2\Sigma^+$ $N'=0$ state.  After a tuned delay, the second laser saturates the $\omega_2$ transition to a selected state in the $\ell=3$ Rydberg series that converges to the NO$^+$ rotational state, $N^+=2$.  Among the states with total angular momentum, $N=1$ accessible from intermediate $N'=0$, only those prepared to occupy this non-penetrating $n_0f(2)$ Rydberg state have predissociative lifetimes long enough to support electron-impact avalanche and evolution to form an ultracold plasma.  

\begin{figure}[h!]
\centering
\includegraphics[width=0.4\textwidth]{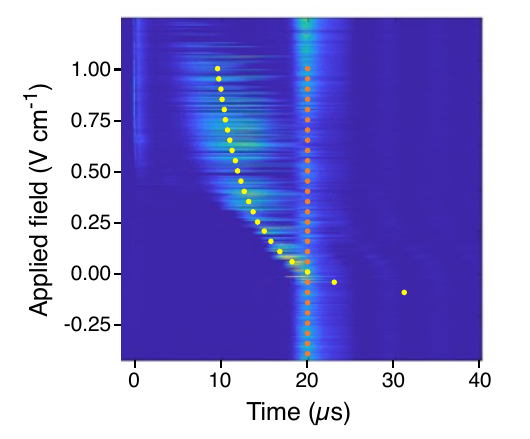}
\caption {\small Shot-by-shot signal produced as the charged volume formed by the  $\omega_1$ plus $\omega_2$ excitation of nitric oxide propagates in a molecular beam to transit the detection grid, G$_2$ at the indicated fields applied between G$_1$ and G$_2$.  (yellow dots) The arrival time of electrons accelerated by the field, calculated figuring the inertial mass of NO$^+$ ions }
\label{fig:scopetrace}
\end{figure}

The moving grid machine offers an informative view of the evolution from state-selected Rydberg gas to plasma as a function of the DC field between G$_1$ and G$_2$.  Figure \ref{fig:scopetrace} stacks oscilloscope traces showing how the plasma signal varies with bias for a G$_2$ position fixed at a distance of 28 mm downstream from the point at which the pair of pulses, $\omega_1 + \omega_2$, illuminate the molecular beam.  Under conditions of +500 mV forward bias, we clearly see an initial signal of prompt electrons coinciding with the firing of $\omega_2$ at $t=0$.  After a flight time of 20 $\mu$s, the illuminated volume, travelling at the propagation velocity of the molecular beam, reaches G$_2$, where it forms a Gaussian electron signal waveform that broadens with an increasing displacement of G$_2$.  

Here, we adjust the G$_1$ - G$_2$ bias continuously from +500 mV (forward) to -500 mV (reverse).  Forward bias produces a broad electron signal waveform that leads the late-peak detection of the illuminated volume.  From the variation of this waveform with G$_1$ - G$_2$ bias, we can conclude that this signal does not arise from free electrons but rather represents the polarization of a loose NO$^+$ - e$^-$ charge distribution (electrons bound by the associated NO$^+$ space charge).  This space charge polarization relaxes to merge with the late-peak illuminated volume signal when we tune the G$_1$ - G$_2$ bias through zero.  Note the evident absence of a polarized electron signal trailing the late peak under conditions of reversed bias.   

The free-electron signal appears with highly variable amplitude.  By contrast, the density of late peak electrons propagating with the illuminated volume remains relatively constant over a wide range of initial Rydberg gas density, $\rho_0$.  

\subsection{Selective field ionization as a gauge of plasma electron binding energy as a function of time }

The SFI spectrum formed by an electrostatic ramp applied at any point before the illuminated volume transits G$_2$ provides a measure of the evolving distribution of plasma electron binding energy as a function of time.  
\begin{figure}[h!]
\centering
         \includegraphics[width=.4 \textwidth]{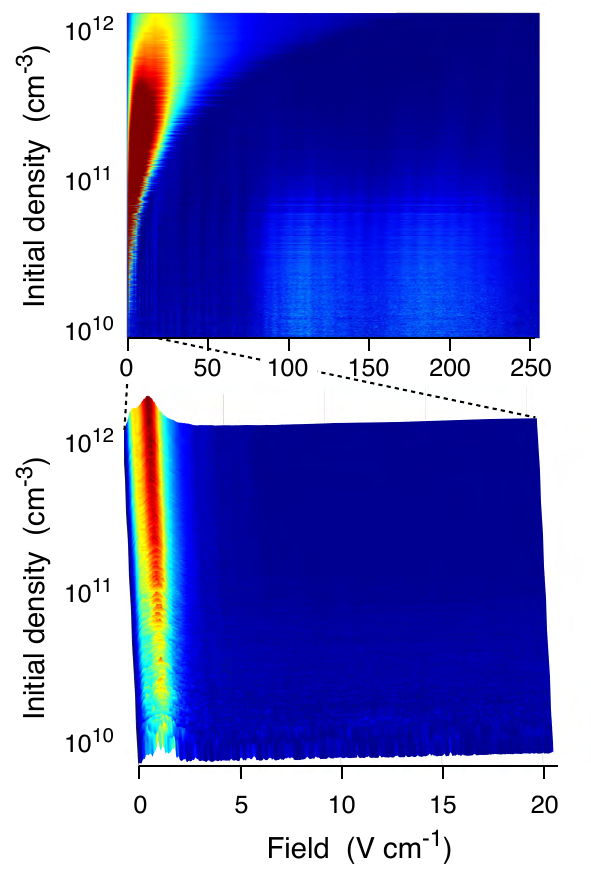}         
     \caption{\small (top) SFI spectrum, formed by  4,000 SFI traces sorted according to the initial density $\rho_0$, for an $nf(2)$ Rydberg with an initial principal quantum number $n_0 = 49$, following a ramp delay, of $t_R=6$ $\mu$s.  (bottom) SFI spectrum of electron binding energies sampled after an evolution of 480 $\mu$s.  } \label{fig:SFI}   
   \end{figure}
Before the start of the ramp, a G$_1$ - G$_2$ potential held to $0 \pm 20$ mV cm$^{-1}$ ensures that the ultracold plasma evolves as a Rydberg gas together with a gas of ions and unpolarized electrons.  A complete account of the electron binding energy spectrum of an ultracold plasma at any point in its evolution considers the contour of SFI traces for a wide interval of initial Rydberg gas densities.   
  
To acquire this complete contour of SFI spectra, we hold the laser powers in a regime of saturated $\omega_1$ and $\omega_2$ transitions and increment the $\omega_1 - \omega_2$ delay in a series of 50 steps, increasing from 0 to 1000 ns.  Over this range, fluorescence of the A $^2\Sigma^+$ state with a lifetime of 196 ns diminishes the effective size of the Gaussian ellipsoid by a factor of 150.  

The avalanche sizes measured by the integrated traces at any given delay fluctuate widely, overlapping with signal intensities observed at many neighbouring initial densities.  To regularize diagnostic aspects of these spectra, we stack traces vertically according to the experimental integrated intensity and then normalize every individual trace.  

During the first 250 ns after $\omega_2$, the electron binding energy of a Rydberg gas with the initial principal quantum number, $n_0$, selected by double resonant excitation, dominates the SFI spectrum \cite{marroquin2024self}.  The characteristic appearance potential pattern reflecting field ionization to form $N^+=0$ and 2 rotational states of NO$^+$ changes to reflect the rising population of plasma electrons.  These free electrons collide with residual Rydberg molecules and shift the Rydberg appearance potential signal from one characteristic of $n_0f(2)$ to that of less-penetrating $\ell$-mixed $n_0\ell(2)$ orbitals, which appears at slightly higher field.

The top frame of Figure \ref{fig:SFI} displays the contour of plasma SFI spectra formed by 4000 $49f(2)$ Rydberg gas volumes of systematically varying initial density obtained for a ramp delay of 6 $\mu$s. Note how Rydberg gases of high initial density avalanche fully, while those of lower density evolve to form a mixture of weakly bound NO$^+$ ions and electrons together with a trace of Rydberg molecules with binding energies in the range of $n_0$, indicated by the dim signal in the bottom right of the SFI spectrum.  An evident absence of electron signal forms a slight wedge at very low SFI potential in the upper left-hand corner of this contour.  At the highest density, field ionization requires a minimum of a few V cm$^{-1}$, corresponding to Coulomb binding energy on the order of 300 GHz.  

For comparison, the lower frame of Figure \ref{fig:SFI} shows the SFI spectrum of the plasma in its arrested state following a flight time of nearly 500 $\mu$s.  The integrated signal per pulse in this long flight-path experiment differs little from the late peak measured at 40 $\mu$s in the moving-grid apparatus.  Note the expanded scale on which we plot the rising amplitude of the electrostatic ramp.  The extraction of electron signal from the arrested plasma state produced by a Rydberg gas of any initial density requires approximately the same minimum applied field of about 500 mV cm$^{-1}$ and nearly every ellipsoidal volume in the initial Rydberg gas size distribution sampled by the sweep of $\omega_1 - \omega_2$ delay yields a plasma state of the same final density.   

\subsection{Plasma decay to a state of invariant integrated density }

\begin{figure}[h!]
\centering
\includegraphics[scale=0.55]{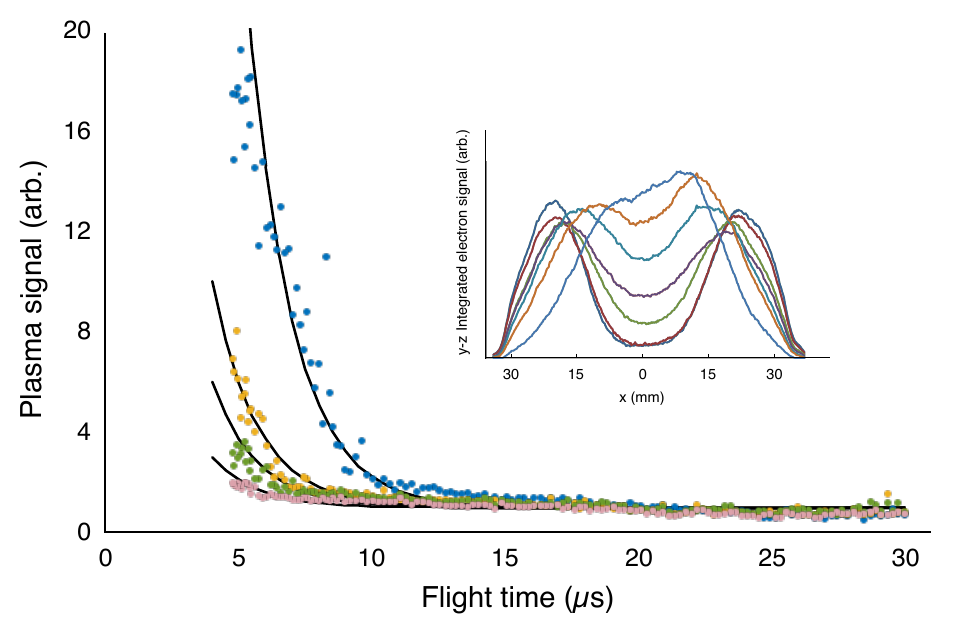}
\caption {\small{Plasma density as a function of the flight time to G$_2$, measured by the area of the integrated plasma signal at zero-field, as in Figure \ref{fig:scopetrace} for $45f(2)$ initial Rydberg gas densities of $4.5 \times 10^{12}, ~9  \times 10^{11}, ~5  \times 10^{11}$ and $2  \times 10^{11}$ cm$^{-3}$, secured by $\omega_1-\omega_2$ delays of 0, 320, 440 and 620 ns.  All curves conform with an exponential time constant of 1.67 $\mu$s.  (inset) Bifurcated traces formed by replicate pairs of $44f(2)$ Rydberg gases after flight times of 500 $\mu$s over a distance of 700 mm for estimated initial densities of $5 \times 10^{11}, ~6.5 \times 10^{11}, ~8 \times 10^{11}$  and $1 \times 10^{11}$ cm$^{-3}$ reading from the smallest to largest separations. }}
\label{fig:density-decay}
\end{figure}

The electron signal waveform detected as the illuminated volume transits G$_2$ measures the early time evolution of the plasma to this evident state of arrested integrated density.  Here, using the moving grid apparatus, we vary the detection time by increasing the beam propagation distance to G$_2$.  Figure \ref{fig:density-decay} traces the integrated late-peak signal as a function of flight time for the ultracold plasma that forms in zero field from a $43f(2)$ Rydberg gas at four different density precisely selected by varing the $\omega_1 - \omega_2$ delay.  

Here, for G$_1$ - G$_2$ distances from 4 to 14 mm (3 to 10 $\mu$s), we see signals that fall from widely different initial amplitudes with the same time constant of roughly 2 $\mu$s.  This time coincides with the observation of widely fluctuating shot-to-shot integrated avalanche sizes.  

By 10 $\mu$s, Rydberg gases of all initial density converge to form an invariant plasma state with a density of $1 \times 10^{11}$ cm$^{-3}$.  Constrained by forces of self-organization, this plasma expands slowly, descending to a Gaussian density of $5 \times 10^{10}$ cm$^{-3}$ after 30 $\mu$s.  After a flight of 500 $\mu$s, arrested ensembles with an approximate density of $3 \times 10^9$ cm$^{-3}$ exhibit the very low electron-binding energy spectrum pictured in the lower frame of Figure \ref{fig:SFI} and form the bifurcated volumes pictured in the inset of Figure \ref{fig:density-decay}.  

\subsection{Effect of radiofrequency excitation on plasma dissipation }

\begin{figure}[h!]
    \centering
        \includegraphics[width= .44 \textwidth]{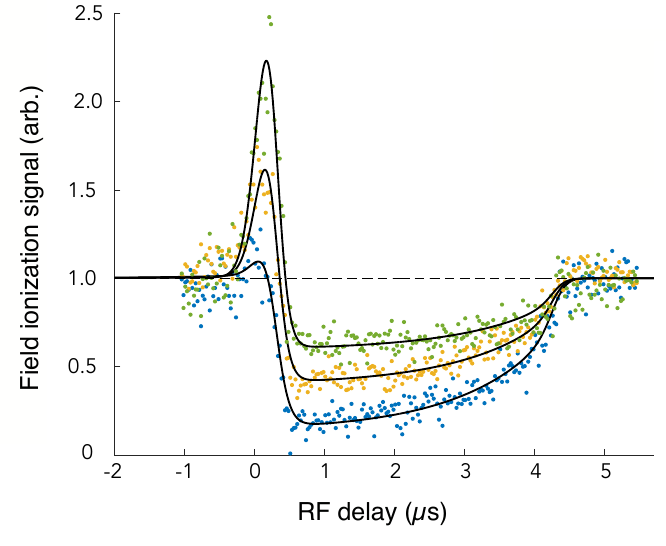}
     \caption{Integrated electron signal obtained by selective field ionization of an ultracold plasma after evolution of 4.8 $\mu$s from a $49f(2)$ Rydberg gas of NO in the presence of a 250 ns  60 MHz pulsed radio frequency field, plotted as a function of RF delay for the three initial densities of approximately $1 \times 10^{12}$,  $7 \times 10^{11}$ and $3 \times 10^{11}$ cm$^{-3}$.  Each curve is normalized according to the signal at -1$\mu$s RF pulse delay when the RF pulse has no effect. The black lines fit these measures of Rydberg amplitude as a function of RF delay to a logistic function as described in reference \cite{wang2020radio}}
    \label{fig:rf-delay}
\end{figure}

Depending on the time of application, a 250 ns $\times$ 400 mV cm$^{-1}$ square-wave pulsed radiofrequency field significantly affects the process of plasma dissipation.  Figure \ref{fig:rf-delay} plots the integrated electron signal extracted by an SFI field ramp that begins 4.8 $\mu$s after $t_{\omega_2}$.   Note that the SFI signal increases for an RF pulse applied during or just after $\omega_2$.  After a delay of 500 ns, an RF field depletes the plasma signal by an amount that depends on the RF delay.  

Note that the depth of this depletion increases with the initial density of the Rydberg gas, as does the slope of its recovery as the RF delay approaches the 4.8 $\mu$s initiation of the field ramp.  At fixed RF amplitude, we see the least depletion and greatest proportional enhancement by far for the Rydberg gas of the lowest initial density.  When applied in the first 250 ns after $\omega_2$, a 400 mV cm$^{-1}$ RF field causes very little enhancement in the SFI signal in the case of a Rydberg gas an initial density of $10^{12}$ cm$^{-3}$ or more.

\subsection{Ultracold plasma high-Rydberg excitation spectrum }

\begin{figure}[h!]
\centering
\includegraphics[scale=0.58]{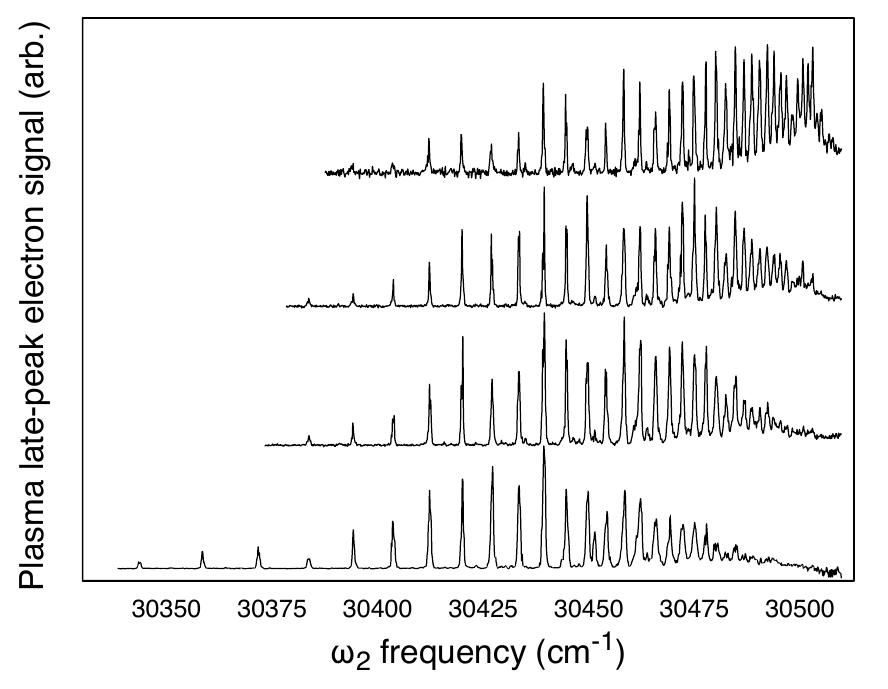}
\caption {\small Long-lived plasma $(\omega_1 + \omega_2)$ excitation spectra obtained for initial densities of $3.0 \times 10^{12}, ~1.0 \times 10^{12}, ~1.5 \times 10^{11}$ and $2.5 \times 10^{10}$ cm$^{-3}$, from bottom to top.}
\label{fig:w2_stack}
\end{figure}

The excitation spectrum of $\omega_2$ transitions, detected as a durable ultracold plasma late peak, spans principal quantum numbers, $n_0$, from 28 to more than 80 in a single Rydberg series assigned to $n_0f(2)$.  These states, composed of Rydberg electrons with orbital angular momentum, $\ell=3$ built upon rotational $N^+ = 2$ NO$^+$ cores, constitute the least penetrating Rydberg states with total angular momentum neglecting spin of $N=1$, accessible by electronic transition from A $^2\Sigma^+$ $N'=0$.  The interval of $n_0f(2)$ states populated by $\omega_2$ absorption varies with the initial density of the Rydberg gas.  Figure \ref{fig:w2_stack} shows the long-lived plasma excitation spectra observed after a flight time of 500 $\mu$s for $\omega_1$ - $\omega_2$ delays of 100, 300, 700 and 1100 ns, yielding initial Rydberg gas densities of $3.0 \times 10^{12}, ~1.0 \times 10^{12}, ~1.5 \times 10^{11}$ and $2.5 \times 10^{10}$ cm$^{-3}$ (3.0, 1.0, 0.15 and 0.025 $\mu$m$^{-3}$).  

\begin{figure}[h]
    \centering
      \includegraphics[width=.45\textwidth]{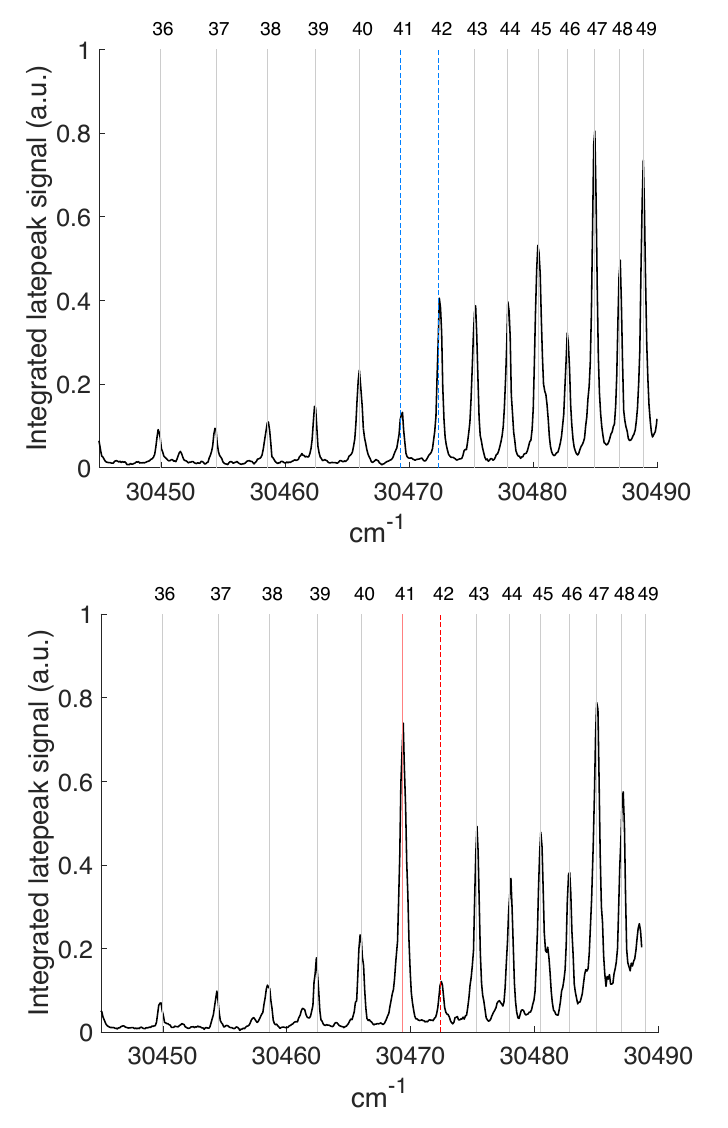}
      \caption{ \small Double-resonant $\omega_2$ spectra of nitric oxide $n_0f(2)$ Rydberg states for $n_0$ from 38 to 49 integrated late peak after a flight time of 20 $\mu$s in the absence (top) and presence (bottom) of a CW mm-wave field tuned to 93.7GHz.}  
     \label{fig:w2}
   \end{figure}

Figure \ref{fig:w2} compares a segment of the double-resonant ultracold plasma excitation spectrum for $\omega_2$ wavelengths scanned from 327.8 nm to 328.4 nm with an initial density of $\sim 1 \times 10^{11}$ cm$^{-3}$.  The upper spectrum shows the late peak signal in the absence of a mm-wave field.  Here, we can assign $n_0f(2)$ features that extend from $n_0=36$ to 49.  Note the relatively low intensity of the $n_0=41$ resonance.  The lower spectrum shows the effect of a 25 $\mu$W {{cm$^{-2}$}} CW mm-wave field at 93.7 GHz on the observed $n_0f(2)$ spectrum. We can see that the mm-wave field dramatically enhances the $41f(2)$ resonance while significantly diminishing the excitation feature associated with the $\omega_2$ transition to $42f(2)$.

\begin{figure}[h!]
    \centering
      \includegraphics[width=.45\textwidth]{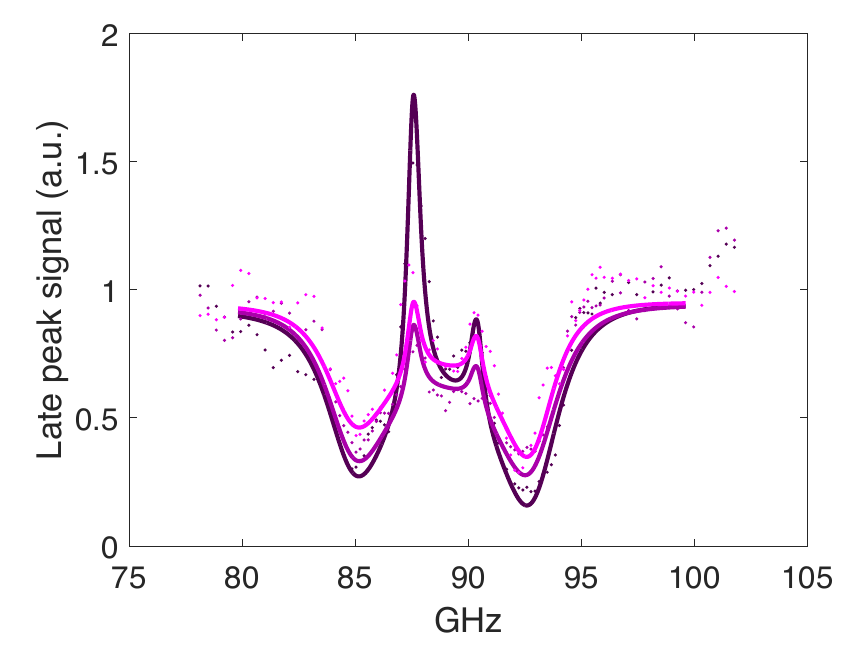}
      \caption{ \small mm-wave excitation spectra observed in the $42f(2)$ resonant ultracold plasma signal detected after a flight time of 20 $\mu$s, for mm-wave frequencies tuned from 80 to 100 GHz.  Features of diminishing depth fit to four-component Fano lineshapes recorded for mm-wave fields triggered to begin at times -1.5 $\mu$s, 3.5 $\mu$s and 8.5 $\mu$s after $\omega_2$ excitation. }  
     \label{fig:mm-spect}
   \end{figure}

Fixing $\omega_2$ on the $42f(2)$ resonance and scanning the frequency of the mm-wave field from 80 to 100 GHz, traces the resonant lineshape of this mm-wave depletion phenomenon shown in Figure \ref{fig:mm-spect}.  Here, we see that mm-wave excitation frequencies of 87.61 and 90.22 GHz enhance the amplitude of the $42f(2)$ resonance.  Scanning this field through the frequency of 93.7 GHz fixed in Figure \ref{fig:w2} forms a spectrum of deep depletion, dipping the electron signal in the ultracold plasma late-peak nearly to zero.  
   
As previously shown \cite{wang2022mm}, each such feature conforms with a four-component Fano lineshape. Each of the two positive-going and two negative-going components fits a distinct interference cross-section formula:  
 \begin{equation}
\sigma(\epsilon)=\frac{(q+\epsilon)^2}{1+\epsilon^2},
\label{eqn:Fano1}
\end{equation}
in which the parameter, $q$ accounts for asymmetry.  The relative absorption cross section, $\sigma(\epsilon)$, varies with detuning, $\epsilon$, defined by:
\begin{equation}
 \epsilon=\frac{\omega_{\rm mm}-\omega_0}{\Gamma/2}
 \label{eqn:Fano2}
 \end{equation}
Here, $\omega_0$ represents the resonant frequency of the transition and $\Gamma$ describes its linewidth.  Peaks decrease rapidly in amplitude for mm-wave fields triggered to begin as little as 1 $\mu$s after $\omega_2$ excitation.  Dips diminish less.
  
     
\begin{table}[h!]
	\centering
	\caption{Fano parameters for $42f(2)$ and $42\ell(2)$ to $(42 \pm 1)g(2)$ (Peak) and $(42 \mp 1)d(2)$ (Dip) Rydberg-Rydberg transitions measured in the late-peak excitation spectrum.    }
	\label{tab:fano-g}
		\begin{tabular}{lrccrcc}
	\toprule
	&    \multicolumn{1}{c} { $\omega_0$ }   &  $\Gamma$   &   $q$   &   \multicolumn{1}{c} { \hspace{0 pt} $\omega_0$ } &  \multicolumn{1}{c} { \hspace{0 pt}  $\Gamma$   }   &   $q$   \\
	 & (GHz) &    (GHz)  &      &    (GHz)&     (GHz)             \\
   			\midrule  
Peak & 87.61   & 0.8    & 15  & 90.22   & 0.8    & -15 \\
Dip & 85.11  & 3 & -4 & 93.07  &  3 & -4  \\
\bottomrule
		\end{tabular}%
\end{table}
 
\subsection{Driven dissipation in the ultracold plasma as a function of mm-wave pulse width and delay}

Previous work has established that mm-wave transitions from initially selected $n_0f(2)$ states to longer-lived $(n_0 \pm 1)g(2)$ states, driven early in the avalanche, act to enhance the conversion of Rydberg gas to plasma \cite{wang2022mm}.  The effectiveness of these $n_0f(2) \rightarrow (n_0 \pm 1)g(2)$ resonances diminish on the 100 ns timescale of $\ell$-mixing early in the avalanche to plasma.  

Depletion, which arises when pulsed mm-wave excitation drives non-penetrating states of high-$\ell$ to highly predissociative $n_0d(2)$ states, persists throughout the 20 $\mu$s time of flight of the illuminated volume from laser intersection with the molecular beam to transmission through the detection grid, G$_2$.  
\begin{figure}[h]
    \centering
      \includegraphics[width=.49\textwidth]{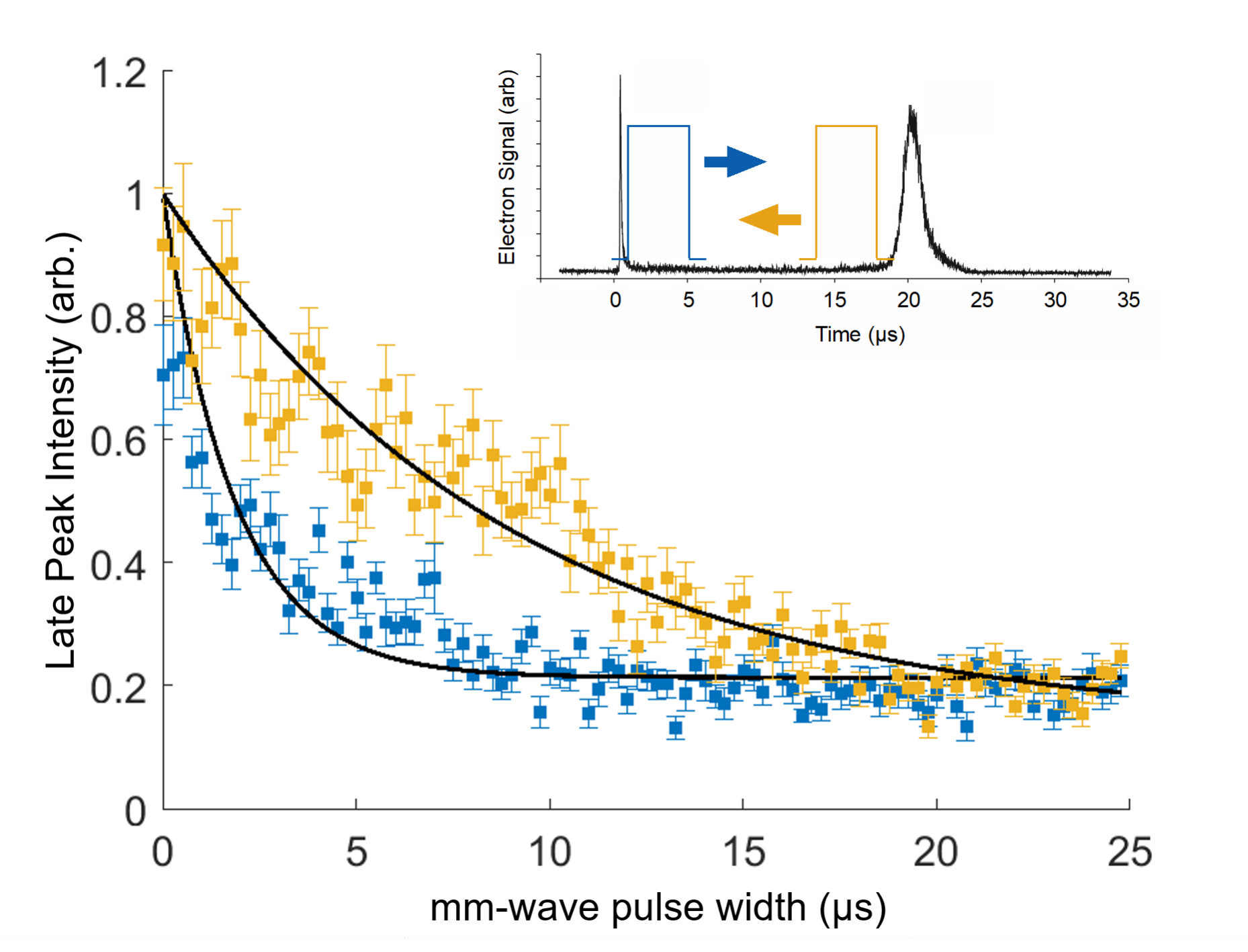}
      \caption{ \small mm-wave driven dissipation of the ultracold plasma that evolves from a $42f(2)$ Rydberg gas, as driven by a mm-wave pulse tuned to the $42\ell(2) \rightarrow 43d(2)$ transition at 93.07 GHz, as a function of pulse width, measured going forward for a pulses that begin at $t_{\omega_2}$ (blue) , and measured backwards from a pulse that ends at $t_{\rm G_2}$ (yellow). }  
     \label{fig:forward_backward}
   \end{figure}
The extent of dissipation depends on the width of the mm-wave pulse, but less on the elapsed time between $t_{\omega_2}$ -- when the $\omega_2$ pulse forms the volume of Rydberg gas entrained in the molecular beam, and $t_{\rm G_2}$ -- when this volume passes through G$_2$ and relays electrons to the MCP detected as the late peak.  
   
   Figure \ref{fig:forward_backward} plots the depth of the dissipation owing to mm-wave driven predissociation of the plasma that evolves from a $42f(2)$ Rydberg gas as a function of the mm-wave pulse width, measured going forward for a pulse tuned to the $42\ell(2) \rightarrow 43d(2)$ transition that begins at $t_{\omega_2}$, and measured backward for a pulse that ends at $t_{\rm G_2}$.  Here, we see that a resonant 5 $\mu$s mm-wave pulse starting at $t_{\omega_2}$ depletes the plasma by about 70 percent.  A similar pulse delayed to start 15 $\mu$s after $t_{\omega_2}$ depletes the plasma by about 40 percent.  Perhaps more telling, 15 $\mu$s mm-wave square wave begun 5 $\mu$s after $t_{\omega_2}$ drives a comparable level of dissipation as the same square wave applied immediately upon plasma formation.

\section{Discussion}

\subsection{Ultracold plasma formation in the field-free limit}

The electron signal waveform observed in the moving-grid spectrometer exhibits a morphology that varies substantially with a positive or negative electrostatic bias applied to G$_2$ over the range of just a few hundred millivolts.  In Figure \ref{fig:scopetrace}, we see that a nominal forward bias causes the appearance of a detected feature that leads the late-peak transit of the illuminated volume through G$_2$.  This leading electron signal waveform grows in amplitude and advances in time with increasing forward bias.  In a limit of a bias reduced to zero-field, this leading signal retreats to join the marked volume of ultracold plasma that propagates with the velocity of the molecular beam.  A reverse bias produces no evident signal waveform trailing the late peak.

We can explain this field-dependent signal as a polarization of free electrons, retarded by a balancing space charge of NO$^+$ ions in the ultracold plasma.  In the 20 $\mu$s between ${\omega_2}$ and the appearance of the marked volume at G$_2$, the ultracold plasma travels 28 mm.  During that time, the electrostatic field between G$_1$ and G$_2$ acts to polarize the electrons that screen NO$^+$ ions in the plasma.  Each electron feels a force directly proportional to the field, $F=qE$, where $q=1.602 \times 10^{-19}$ Coulombs refers to its charge.  Under the constraints of quasi-neutrality assumed in these weak-field conditions, the displacement of an electron requires the associated motion of an ion.  Thus, polarization, in this case, is manifested by an acceleration, $a=F/m$, where $m$ is the mass of an ion.  For NO$^+$, $m$ equals $1.66054 \times 10^{-27}$ kg.  Considering a distance, $s=28$ mm and the velocity of the marked volume in the molecular beam, $u=1400$ ms$^{-1}$, the time of flight of the polarized volume must satisfy the kinematic equation:
\begin{equation}
s = ut + \frac{1}{2}at^2.
\label{eqn:kinematic}
\end{equation}
Solving the equation for $t$, we find arrival times in an accelerating field shortened according to a locus of times indicated by dots in Figure \ref{fig:scopetrace}.  Note how shortening of the flight time diminishes the degree to which higher forward bias can increase this polarization.  

Reverse bias acts in the opposite way.  A retarding potential as small as 100 mV cm$^{-1}$ reduces the velocity at G$_2$ to 900 m s$^{-1}$.  This lengthens the arrival time of the polarized charge distribution at the detection plane to more than 30 $\mu$s, amplifying its deceleration.  The asymmetry of electron polarization by forward versus reverse bias allows us to easily recognize a precise condition of homogeneous zero-field.  

\subsection{Selective field ionization and the initial dynamics of molecular ultracold plasma evolution}

The SFI spectrum of a state-selected NO Rydberg gas measured with no ramp-field delay after $\omega_2$ excitation produces a broad Gaussian distribution of single-shot integrated SFI intensities, in which the binding energy spectrum largely consists of a simple pair of features reflecting the field-ionization threshold of the selected $49f(2)$ high Rydberg state with respect to the $N^+=0$ and 2 rotational states of the NO$^+$ ion \cite{Jones:2008fu,marroquin2024self}.  

By contrast, after a ramp delay of no more than a few $\mu$s, the SFI spectrum shows little, if any, signature of the initial Rydberg gas.  As shown in Figure \ref{fig:SFI}, electrons accounting for most of the signal in the 6 $\mu$s delayed SFI spectrum diabatically bind to $N^+=2$ rotational-state NO$^+$ ions by a vertical energy of no more than 540 GHz.  A light-blue region in the lower right of Figure \ref{fig:SFI} signifies a small residue of Rydberg molecules with the initially selected principal quantum number.  The shift of this signal to a slightly higher appearance potential in the first 200 ns of evolution directly signals the effect of $\ell$-mixing electron collisions that drive residual Rydberg molecules to states of high electron orbital angular momentum.  Hundreds of measurements such as these at each of 50 selected $\omega_1-\omega_2$  delays from 0 to 1000 ns yield wide distributions of integrated intensities, reflecting individually and in aggregate, a power law avalanche size distribution suggestive of scale invariance and a self-organized critical phase \cite{marroquin2024self}. 

This evolution to smaller electron binding energy continues as the plasma proceeds on a flight of nearly 500 $\mu$s to enter the extraction region of a downstream SFI spectrometer.  Here, we find that Rydberg gases spanning two orders of magnitude in initial density decay to form a uniform sub-critical final state characterized by a vertical electron binding energy no greater than 150 GHz and a density calibrated in previous experiments to be approximately $3 \times 10^9$ cm$^{-3}$ \cite{marroquin2024self}.  

\subsection{An emergent state of invariant plasma density}

The electron signal as a function of G$_2$ position, as pictured in Figure \ref{fig:density-decay}, provides a direct means to follow the loss of plasma density during the initial 30 $\mu$s of evolution.   Rydberg gases of all densities decay with the same time constant of about 600 ns.  SFI spectra recorded during this time exhibit power-law avalanche size distributions suggestive of self-organization to a critical state \cite{marroquin2024self}.  The plasmas traced here evolve from Rydberg gases, prepared to span more than a factor of 20 in initial density.  Yet, at all scales, these volumes decay at the same rate, and after 10 $\mu$s, we find the G$_2$-sampled signal relaxes to a single constant, integrated amplitude.   The width of the plasma waveform transiting G$_2$ gauges the degree to which the plasma expands at any flight time \cite{sadeghi2012dissociative,Haenel2017}.  Accordingly, the asymptotic integrated signal in Figure \ref{fig:density-decay} yields an estimated density of $1.2 \times 10^{10}$ cm$^{-3}$ after a flight of 20 $\mu$s.  

\subsection{High Rydberg spectroscopy as a gauge of ultracold plasma dissipation}

\subsubsection{Penning ionization and predissociation shape the $\omega_2$ spectrum}

Transitions involving intermediate resonance with A $^2\Sigma^{+} $ $N^{'}=0$ must terminate in states of $N=1$.  The penetrating $s$, $p$ and $d$ Rydberg states of nitric oxide all rapidly predissociate.  Among accessible states, only the $f$ series converging to $N^+=2$, $n_0f(2)$, forms a long-lived molecular ultracold plasma.  For a Rydberg gas prepared with any principal quantum number, prompt Rydberg-Rydberg Penning ionization of excited NO molecules in the leading edge of the nearest neighbour distribution determines the initial [NO$^+$] and [e$^-$] density:  
\begin{equation}
{\rm{NO^*({\it {n_0}}) +{\rm{NO^*({\it {n_0})}}} \rightarrow NO^+ + e^- +  N(^4S) +O(^3P)}}
\label{eqn:penning}
\end{equation}

The Rydberg orbital radius, which increases as $n_0^2$, sets the critical nearest-neighbour distance, $r_c$, for prompt Penning ionization.  We can find the fraction of molecules that fall within that distance for any initial density by reference to the Erlang distribution \cite{Torquato.1990}.  For any principal quantum number, $n_0$ and Rydberg gas density, $\rho_0$, the fraction of NO$^*$ molecules consumed by prompt Penning interactions equals the integral of this distribution from $r = 0$ to $r_c$, expressed in closed form by:
\begin{equation}
\rho_{\rm NO^+}(\rho_0,n_0) = \frac{\rho_0}{2}(1-\mathrm{e}^{-\frac{4\pi}{3}\rho_0 r_c^3}) 
\label{eqn:Erlang}
\end{equation}
Here, we assume that the perturbation associated with electronic energy transfer causes the Penning partner to sample a manifold of predissociative crossings that lead efficiently to neutral products.  

Evolution to plasma and relaxation follows, forming a signal after 500 $\mu$s that we record as an $\omega_2$ excitation spectrum of the $n_0f(2)$ series.  In Figure \ref{fig:w2_stack}, we see that this spectrum extends to higher principal quantum numbers with decreasing Rydberg gas density.   This trend suggests that $n_0f(2)$ Rydberg states with large Penning ionization cross sections form stable molecular ultracold plasmas only under conditions of low density.  

For all densities, conditions favour the formation of a long-lived plasma state for $n_0$ and $\rho_0$ such that Penning ionization converts about half the NO$^*$ Rydberg population to NO$^+$ ions and companion dissociation products.  Reducing the initial density raises the optimum initial principal quantum number, balancing the population of residual Rydberg molecules with Penning ions formed at larger average NO$^*$-NO$^*$ distances. 

\subsubsection{Electron-Rydberg collision dynamics and $\omega_2$ feature intensities}

With a nominal orbital angular momentum $\ell=3$, the $n_0f(2)$ states form the least-penetrating NO Rydberg series accessible by one-photon absorption from the $N'=0$ level of the A-state.  Yet, the natural predissociation of $nf$ states proceeds on a nanosecond timescale.  Conventional measurements often sample the characteristics of resonances narrowed by virtue of Stark mixing with less-penetrating states of higher $\ell$~\cite{vrakking1995life}.  

The lifetime-lengthening utility of such interactions is made evident in molecular ultracold plasma dynamics by the early time enhancement of plasma formation in a radiofrequency field.  Here, the RF field mobilizes electrons, accelerating e$^- + {\rm NO^*}$ collisional $\ell$-mixing \cite{wang2020radio}.  

As evident in Figure \ref{fig:rf-delay}, the advantage of RF mixing disappears on a 100 ns timescale as newly formed plasma electrons collisionally $\ell$-mix residual NO Rydberg molecules in the evolving plasma.  This electron-collisional $\ell$-mixing combines with the spontaneous predissociation of ${\rm NO^*}$ molecules of low-$\ell$ to create a gapped ensemble of high-$\ell$ molecules, forming a key component of a long-lived prethermal state.   

We subsequently find that applying an RF field to the stabilized population of high-$\ell$ molecules decreases plasma density by bridging the angular momentum gap, causing the stable ensemble to develop low-$\ell$ character and predissociate. The density dependence of these effects proves an electron collisional mechanism for both lifetime lengthening and shortening consequences of $\ell$-mixing.  

\subsubsection{mm-wave spectroscopy:  Electromagnetically induced plasma conservation and dissipation}

In Figure \ref{fig:w2}, we see an example in which a mm-wave field tuned to 93.7 GHz dramatically enhances the intensity of the plasma signal associated with an $\omega_2$ tuned to prepare an initial $41f(2)$ Rydberg gas and suppresses the plasma formed by the adjacent resonance with $41f(2)$.  We attribute enhancement to the population transferred by the mm-wave field from $41f(2)$ to longer-lived $42g(2)$.  A field of this frequency also resonantly transfers population from $42f(2)$, or nearly iso-energetic $42\ell(2)$, to strongly predissociated $43d(2)$.  

Similar transitions are evident in the plasma excitation spectrum associated with a time series of mm-wave scans in Figure \ref{fig:mm-spect} addressing a Rydberg gas prepared by $\omega_1+\omega_2$ double resonance to initially occupy the $42f(2)$ state of NO.  Here, a mm-wave field applied immediately after $\omega_2$ enhances the plasma signal when resonant with transitions from $42f(2)$ to $(42\pm 1)g(2)$ while forming broader deep depletion resonances associated with transitions to $(42 \mp 1)d(2)$.  

On the sub-microsecond timescale of electron-collisional $\ell$-mixing, the enhancement resonances associated with $42f(2)$-state stabilization disappear.  However, the dips remain, even in the regime of $\ell$-mixing.  This reflects the fact that nearly isoenergetic mm-wave transitions from $42\ell(2)$ to $(42 \mp 1)d(2)$ can also support depletion.  This depletion persists throughout the observation window of 20 $\mu$s selected in Figure \ref{fig:forward_backward} despite the fact that delayed SFI spectra show that avalanche drives most of the $n_0f(2)$ molecules to states of very low electron binding energy out of resonance with the mm-wave field in 2 $\mu$s or less.  

The mm-wave field requires time to drive plasma dissipation.  As shown in Figure \ref{fig:forward_backward}, a 93.7 GHz square wave pulse begun immediately after $\omega_2$ must be applied for 10 $\mu$s to maximally deplete the plasma signal.  Reasonably, driving the $n_0\ell(2) \rightarrow (42 \mp 1)d(2)$ transition has a greater effect at the beginning of plasma evolution when the $n_0\ell(2)$ population is high.  But, remarkably, a 5 $\mu$s mm-wave pulse begun after a delay of 15 $\mu$s is only a factor of 2 less effective, even though the $n_0\ell(2)$ population at that point is undetectable by SFI.  

We can understand that these electromagnetic transitions convert some small fraction of the Rydberg molecules in the ensemble to states of low-$\ell$.  Rydberg-Rydberg coupling must spread the dissipative character of these penetrating states in the ensemble, bridging the angular momentum gap to the ${\rm {N(^4S) +O(^3P)}}$ predissociation continuum.  This redistribution evidently broadens the response and deepens depletion by preventing hole burning despite a negligibly small number of bright states at any mm-wave frequency.  The following analysis seeks to illustrate this conceptually in terms of a primitive model for a disordered open quantum system.

\section{Theoretical approach}

The results described above establish that a state-selected molecular Rydberg gas of nitric oxide evolves naturally to a quenched state of canonical ultracold plasma density.  A low-amplitude RF field mobilizes electron collisions, breaking this prethermalized state.  A very similar deep depletion arises with the promotion of a sparse population of residual Rydberg molecules from non-penetrating $n_0 \ell (2)$ levels to predissociative $(n_0 \pm 1) d(2)$ states.  Both effects signal a profound effect of dissipation on prethermal plasma stability.  

Building on our conceptual model for arrested relaxation in an ultracold plasma state of randomly interacting dipoles of random energies~\cite{Sous2018,Sous2019}, we see mm-wave driving as a direct means for adding dissipative states to the collisionless ensemble.  The rapid predissociation of these states serves to bridge the closed quantum system to a thermal continuum of free ${\rm{N (^4S) + O (^3P)}}$ atoms, in effect creating an open quantum system that is weakly coupled to a reservoir of the free-atom environment. Working within the standard Born-Markov approximation, we model these dynamics using a Lindblad master equation~\cite{oxford} that incorporates disorder as well as localized dissipation that allows the system to incoherently relax to a non-equilibrium steady-state. Here, we provide numerical preliminary results on this system which establishes it as a toy model within which to explore the rich dynamics observed in the experiment; a thorough theoretical analysis of this model will be presented in a separate forthcoming work. 

\subsection{Open quantum systems within the Lindblad formalism}

The density matrix $\rho_{T}$ for a closed quantum system undergoes unitary evolution governed by the von Neumann equation:
\begin{equation}
    \dot{\rho}_T=-\frac{i}{\hbar}[H,\rho_T] \, ,
    \label{eqn:VN}
\end{equation}
which is equivalent to the Schr\"odinger equation (in the Schr\"odinger picture). For a system (S) that is in contact with some environment (E), the combined evolution follows the von Neumann equation~\eqref{eqn:VN} while the dynamics of the system alone is generically non-unitary. Specifically, tracing out the environmental degrees of freedom, one obtains the reduced density matrix $\rho \equiv \rho_{S} =\Tr_{E}(\rho_{T})$ for the system. The Gorini Kossakowski Sudarshan Lindblad (GKSL) formulation~\cite{lindblad1976,gorini1976completely} provides a general trace-preserving and completely positive expression for the evolution of $\rho$, referred to as the Lindblad master equation (see Ref.~\cite{AJP} for a review):
\begin{equation}
    \dot{\rho}=\mathcal{L}(\rho)= -\frac{i}{\hbar}[H,\rho]+\sum_j\gamma_j(L_j\rho L_j^+-\frac{1}{2}\{L_j^+L_j,    \rho \}) 
    \label{eqn:Lb}
\end{equation}
where $L_j$ represents a ``jump" operator that describes the coupling between the system and the environment, with the corresponding coupling rate $\gamma_j$ (with $\gamma_j \geq 0 \, \forall \, j$). Starting from a microscopic description of the combined system and environment, the Lindblad equation Eq.~\eqref{eqn:Lb} can be derived within the weak-coupling Born-Markov approximations along with the rotating-wave approximation~\cite{oxford,weimer2021simulation}, all of which we assume here. In particular, here we will treat the Lindblad equation as a useful phenomenological description of our system that captures the correct qualitative features of the observed relaxation dynamics. 

In what follows, it will prove useful to map the $d$-dimensional Hilbert space to the $d^2$ dimensional Fock-Liouville space, in which the system density matrix is represented as a vector $\ket{\rho}$ and operators are represented by the Liouville super-operators. We can then numerically simulate the dynamics for the density matrix elements (given by Eq.~\eqref{eqn:Lb}) in terms of $d^2$ coupled rate equations using the QuTiP toolbox~\cite{qutip1,qutip2}.

\subsection{Long-range interacting 1D Spin-$\frac{1}{2}$ XY chain}

\begin{figure}[t]
    \centering
    \includegraphics[width=0.5\textwidth]{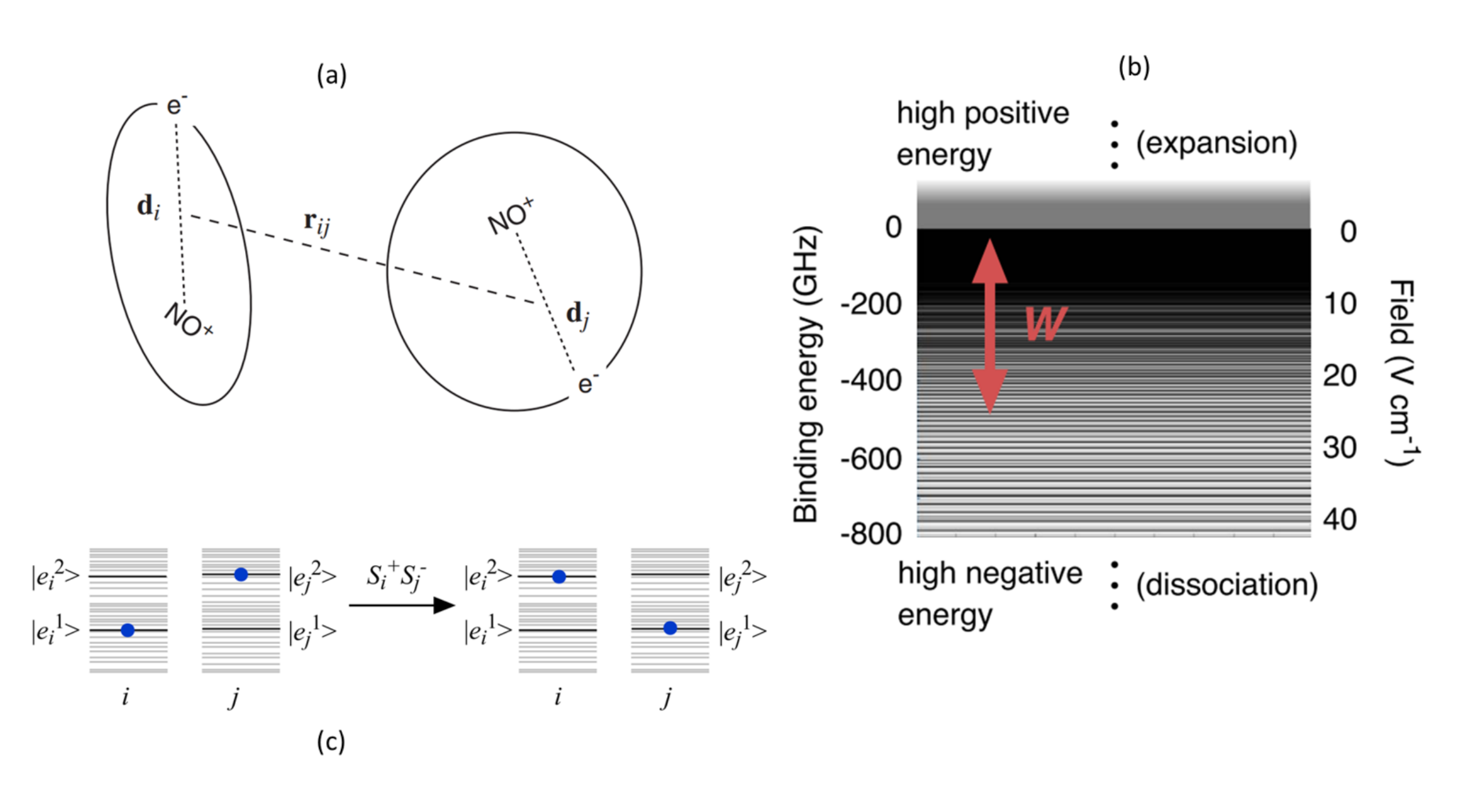}
    \caption{(a) Schematic diagram of NO$^+$ core ions paired with electrons to form interacting dipoles  $\textbf{d}_i$ and $\textbf{d}_j$ separated by $\textbf{r}_{ij}=\textbf{r}_{i}-\textbf{r}_{j}$. The interaction potential, to lowest order, is $V^{dd}_{ij}=[\textbf{d}_i\cdot\textbf{d}_j-3(\textbf{d}_i\cdot \textbf{r}_{ij})(\textbf{d}_j\cdot \textbf{r}_{ij})]/ \textbf{r}_{ij}^3$.  (b) Illustration of measured distribution of binding energies, $W$, defined by SFI spectra. (c) Schematic diagram of the spin-exchange interaction $\hat{S}_i^+ \hat{S}_j^-$ of two Rydberg molecules in the two-level (spin-half) approximation (figures taken from Ref.~\cite{Sous2019}).}
    \label{fig:MBL-model}
\end{figure}

We consider an effective many-body Hamiltonian governed by pair-wise dipolar interactions between ion-electron pairs. In our experimental landscape, quenching gives rise to a wide distribution of random potentials. In this dipole-dipole coupling limit, we can represent pairwise excitations by spins with energy $\epsilon_i$ and exchange interactions governed by a disordered and long-range interacting XY Hamiltonian (see Fig.~\ref{fig:MBL-model}):    
\beq
\label{eqn:ham}
H = \sum_i \epsilon_i \hat{S}_i^z + \sum_{i,j} J_{ij} \left(\hat{S}_i^+ \hat{S}_j^- + \text{h.c.} \right),   
\eeq
where ${\hat{S}^\alpha}=\hbar \hat{\sigma}^\alpha/2$ denotes a spin-$S$ operator for which $\hat{\sigma}^\alpha$ is the corresponding Pauli matrix (for $\alpha =x,y,z$) and $\hat{S^{\pm}}=\hat{S}^x \pm i\hat{S}^y$ are the spin raising and lowering operators.

The first term in Eq.~\eqref{eqn:ham} describes the diagonal disorder that stems from the randomly distributed on-site energy of any particular dipole with disorder strength $W$, for which $\epsilon_i \in [0,W]$. The representative selective field ionization spectrum gauges a $W \sim$500 GHz for the quenched ultracold plasma. The $J_{ij}=\bar{J}/r_{ij}^3$ term determines coupling strength of the spin flip-flop interactions that contribute to the off-diagonal disorder amplitude. $\bar{J}$ varies as $\bar{J} \propto |\textbf{d}_i| |\textbf{d}_j|$, where $\textbf{d}_i$ and $\textbf{d}_j$ are the dipole moments induced by the local ion-electron pairs. Over the present range of $W$, a simple pair of dipoles formed by $s$ and $p$ Rydbergs with the same $n$ has a coupling strength $\bar{J} \sim$ 75 GHz $\mu m^3$ \cite{zoubi2015van}.
\begin{figure}[t]
    \centering
    \includegraphics[width=0.5\textwidth]{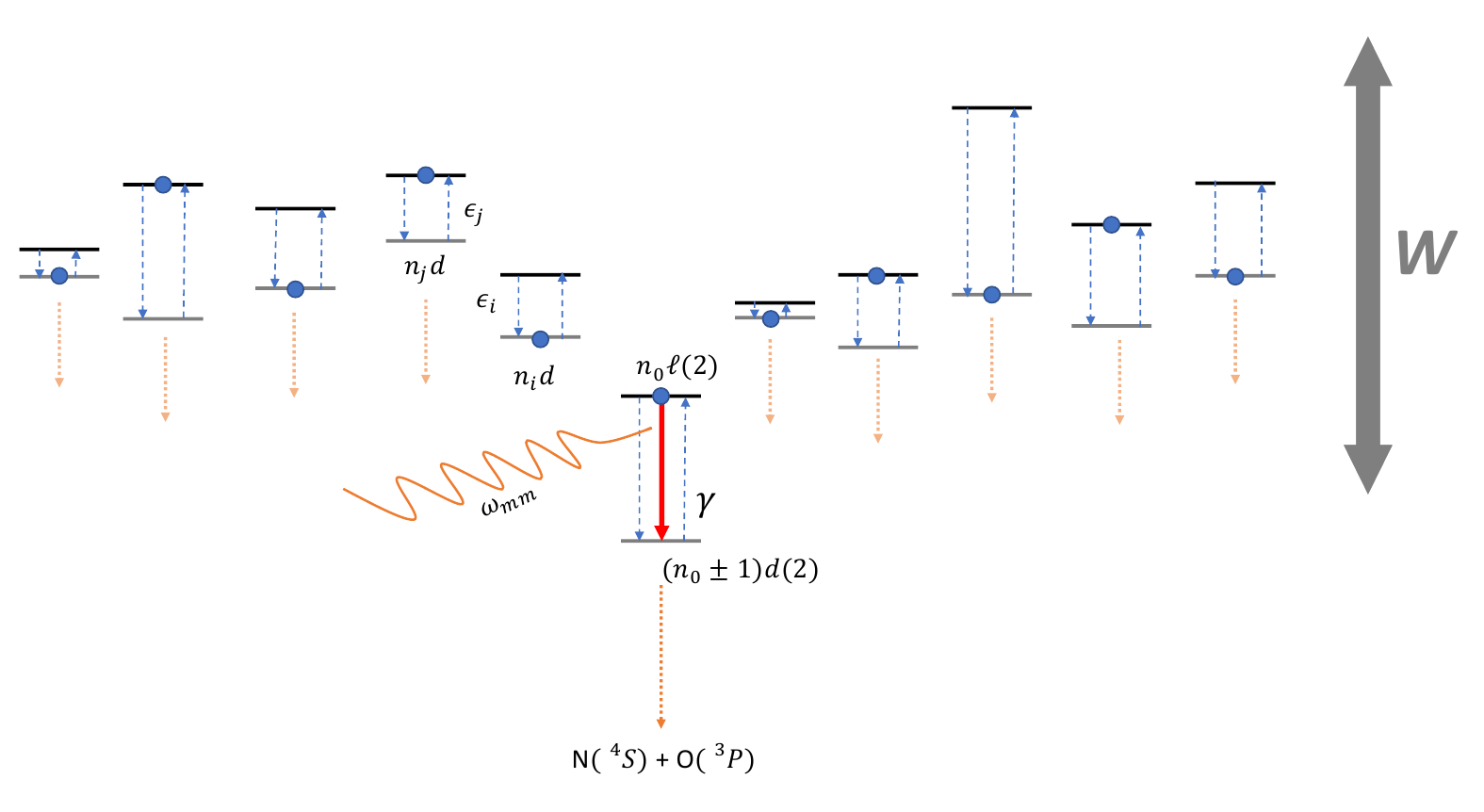}
    \caption{Schematic representation of the mm-wave field driven plasma as a 1D spin-$1/2$ chain with on-site energy disorder and subject to local dissipation.}
    \label{fig:spin-diagram}
\end{figure}

Here, we neglect the van der Waals interaction term $\sum_{i,j} U_{ij} \hat{S}_i^z \hat{S}_j^z$ and make two further simplifications: first, we consider a one-dimensional (1D) system with $L$ sites (with lattice spacing $a=1$) and second, we restrict ourselves to a two-dimensional on-site Hilbert space i.e., we only consider effective spin-$1/2$ degrees of freedom which describe two internal dipolar levels that are well-isolated from the remainder. Fig.~\ref{fig:spin-diagram} schematically depicts this picture of the Rydberg plasma as a 1D model of spin-$1/2$'s with on-site disorder, on which the mm-wave field generates dissipation localized at a single site. Here, we expect that the local dissipation occurs in the residual $n_0\ell(2)$ Rydberg states that occupy less than ten percent of the total population in the arrested NO ultracold plasma system. The mm-wave field then populates the $(n_0\pm1)d(2)$ state which immediately decays into the system of N($^4$S) + O($^3$P) atoms. The long-range dipolar interaction causes the non-resonant states, which occupy a large population and span a wide range of binding energies $W$, to transition into accessible dissociative states. 

Despite these simplifications, in the following we focus on the distributed effects of local dissipation---modeled by the local jump operator $L_j = S_j^-$ acting on site $j$ in Eq.~\eqref{eqn:Lb}---on this many-body system and show that it qualitatively reproduces the observed relaxation dynamics. We note that this relatively simple 1D model cannot be mapped onto a quadratic free-fermion model via a Jordan-Wigner transformation unless the jump operator acts on only the first ($j=0$) and/or last $j=L-1$ site of the chain (with open boundary conditions)~\cite{prosen2008general,prosen2008qpt,dutta2020}; since we do not restrict to this case, the model studied here constitutes a genuinely interacting open many-body system. Note also that in contrast to local dephasing ($L_j = S_j^z$), which preserves the strong U(1) symmetry of the closed system dynamics, the local dissipation channel considered here only has a weak U(1) symmetry~\cite{albert2014}, such that the total magnetization is no longer conserved under the open system dynamics.

\subsection{Numerical results}

We now consider the dynamics of an initially pure state under the Lindblad master equation Eq.~\eqref{eqn:Lb}, with coherent dynamics generated by the Hamiltonian in Eq.~\eqref{eqn:ham} and with incoherent relaxation processes given by the jump operator $L_j = S_j^-$ acting only on a single site $j$. We consider a chain of length $L=11$ and take the Ne\'el state $\ket{\psi} = \ket{\uparrow \downarrow \dots \downarrow \uparrow}$ as the initial state, such that the central site is initially in the $\ket{\uparrow}$ state. We use open boundary conditions throughout and work in units where $\hbar=1$, with the competition between the long-range interactions $J_{ij}= \bar{J}/r_{ij}^3$, disorder $W$, and dissipation $\gamma_i = \gamma \delta_{i,j}$ strengths setting the relaxation dynamics. To study this interplay, we numerically investigate the time-evolution of the local magnetization $\langle S_i^z \rangle$ under the open system dynamics.  

\subsubsection{Dissipation without disorder: W=0}

\begin{figure}[t]
    \centering
    \includegraphics[width=0.5\textwidth]{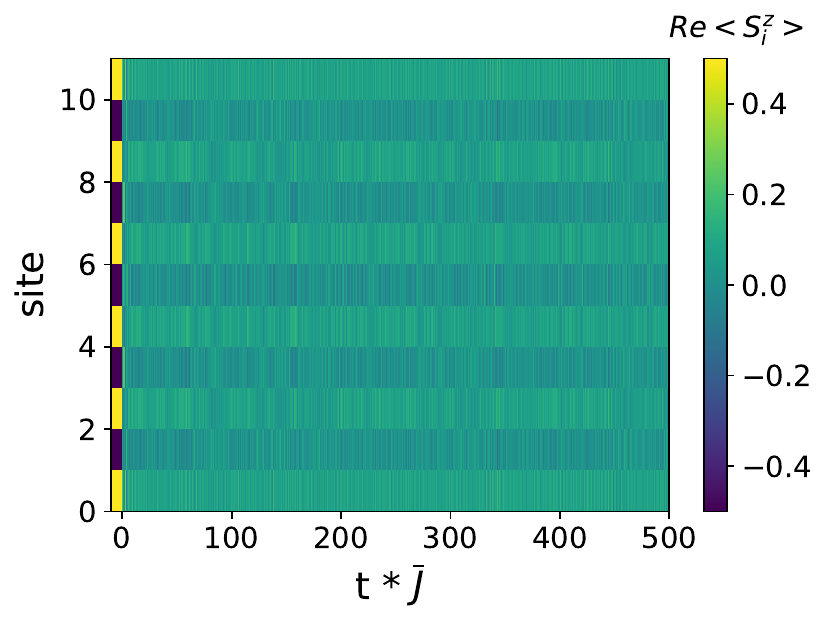}
    \caption{Time-evolution $\langle S_i^z(t) \rangle$ in a closed system of $L = 11$ spins with unit coupling ($\bar{J}=1$), zero disorder ($W=0$), and zero dissipation ($\gamma=0$).}
    \label{fig:W=0 g=0}
\end{figure}
As a baseline, we first consider the closed system ($\gamma=0$) evolution of the clean system ($W = 0$), which conserves the total spin $\sum_{j} S_j^z$ due to the U(1) spin-rotation symmetry. Fig.~\ref{fig:W=0 g=0} shows the coherent dynamics starting from the Ne\'el state.

\begin{figure}[t]
    \centering
    \includegraphics[width=0.5\textwidth]{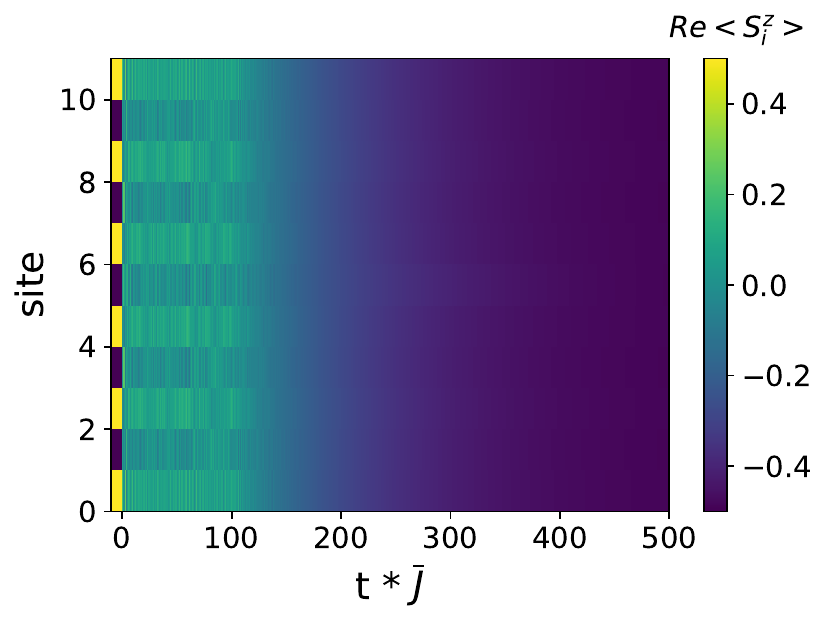}    
    \caption{Time-evolution $\langle S_i^z(t) \rangle$ of the same system as in Fig.~\ref{fig:W=0 g=0}. The system is allowed to evolve coherently up to time $t*\bar{J}=100$ at which point we switch on dissipation that acts locally at the central site $j=5$ with strength $\gamma = 0.1$.}
    \label{fig:W=0 g=0.1}
\end{figure}

To study the effect of dissipation, we first allow the system to evolve coherently for a time $t*\bar{J} = 100$ and then switch on local dissipation (encoded in the jump operator $L_j = S_j^-$) that acts only at a single site. In the absence of any disorder, even a small amount of local dissipation $\gamma \gtrsim 0.1$ leads to rapid relaxation across the entire system, consistent with a spectrum of delocalized eigenstates in the clean long-range interacting XY chain. This is shown in Fig.~\ref{fig:W=0 g=0.1} for dissipation of strength $\gamma = 0.1$ applied to the central site, where we clearly observe that the local magnetization $\langle S_i^z \rangle$ decays to $-1/2$ on each site at the same rate once dissipation is added.

\subsubsection{Dissipation with strong disorder: W=10}

\begin{figure}[t]
    \centering
    \includegraphics[width=0.45\textwidth]{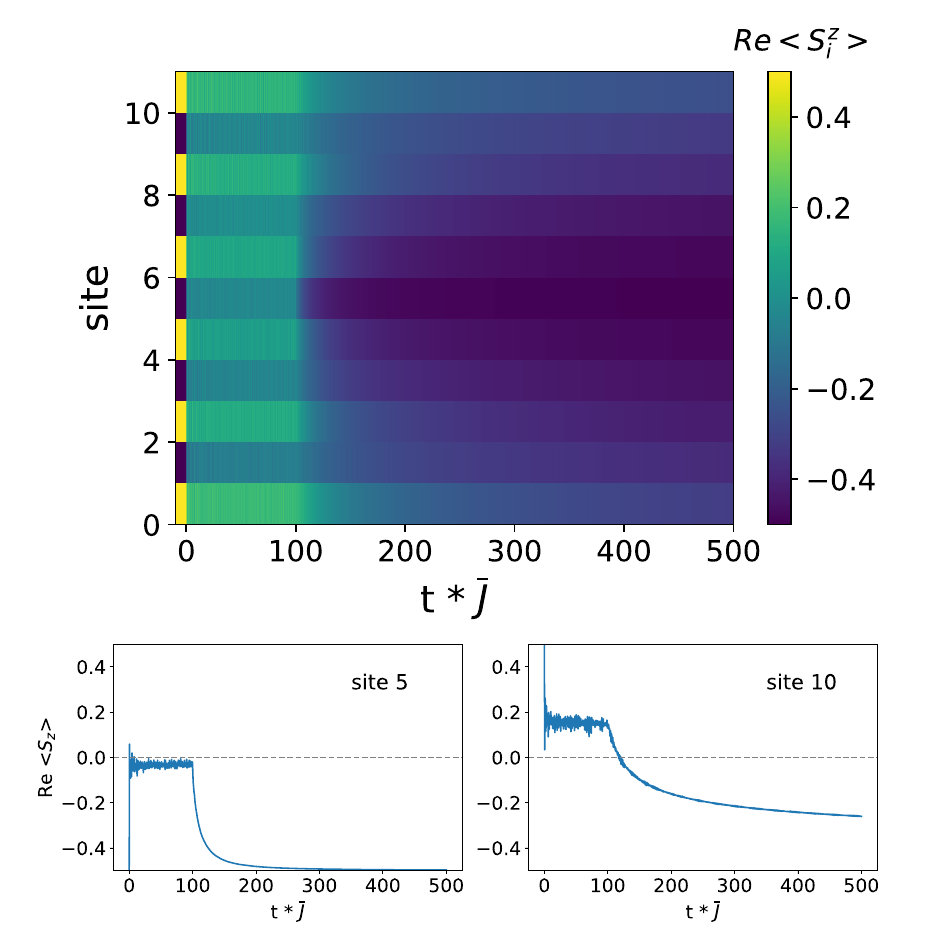}
    \caption{Above: Time-evolution $\langle S_i^z(t) \rangle$ in a system of $L=11$ spins with unit coupling ($\bar{J}=1$), strong disorder ($W=10$), and moderate dissipation ($\gamma_j =0.5$) applied at $t*\bar{J}=100$ to the central site $j=5$. Below: Local spin dynamics at the dissipative site $j=5$ (left) and at a boundary site $j=10$ (right) clearly display distinct relaxation time-scales.}
    \label{fig:W10 center}
\end{figure}

\begin{figure}[t]
    \centering
    \includegraphics[width=0.46\textwidth]{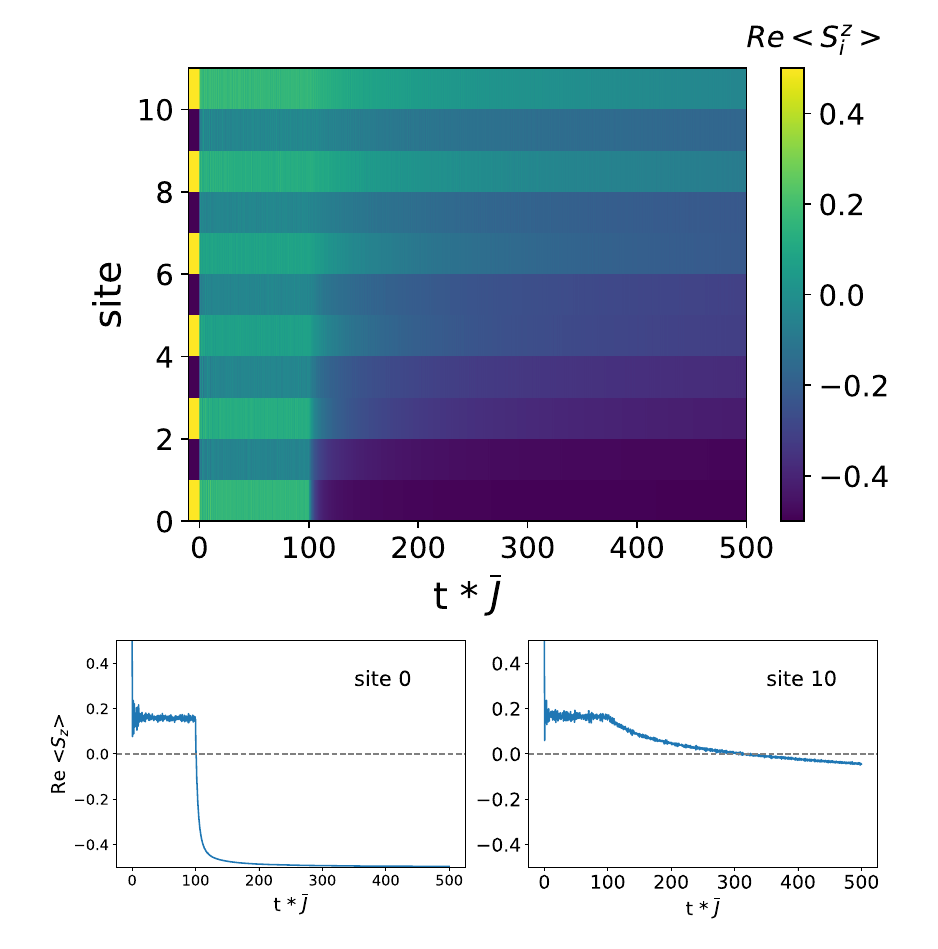}
    \caption{Above: Time-evolution $\langle S_i^z(t) \rangle$ in a system of $L=11$ spins with unit coupling ($\bar{J}=1$), strong disorder ($W=10$), and moderate dissipation ($\gamma_j =0.5$) applied at $t*\bar{J}=100$ to a boundary site $j=0$. Below: Local spin dynamics at the dissipative site $j=0$ (left) and on the other boundary $j=10$ (right) clearly display distinct relaxation time-scales.}
    \label{fig:W10 bound}
\end{figure}

We now investigate the interplay between local dissipation and disorder in the presence of the long-range XY interactions. In particular, we study the case where the on-site disorder is the largest energy scale in the system and set $W = 10$, $\bar{J} = 1$, and consider weak dissipation $\gamma = 0.5$ acting on either the boundary ($j=0$) or central ($j=5$) site of an $L=11$ site chain. We also perform disorder-averaging over twenty different disorder configurations $\{\epsilon_j\}$, where $\epsilon_j \in [0,W] \, \forall \, j$. As before, we allow the system--initialized in the Ne\'el state--to first evolve coherently from time $t=0$ to $t = t*\bar{J}$, at which point the local dissipation is switched on.

Since we are working in the strongly disordered regime, we expect that while the dissipative site will rapidly relax to the spin-$\downarrow$ state, localization will protect the remaining sites from decoherence for longer time-scales. Note however that even exponentially localized states will relax at long times due to the non-trivial overlap between the eigenstate localized at the dissipative site and those localized at distant sites. Thus, only systems with strictly localized eigenstates are expected to remain localized in the long-time steady state, while systems with power-law or exponentially localized states will eventually reach $\langle S_i^z(t) \rangle = -1/2$ in the limit that $t \to \infty$, albeit at a rate that is suppressed compared to the clean case. 

Figs.~\ref{fig:W10 center} and~\ref{fig:W10 bound} illustrate the relaxation dynamics for dissipation acting on the central ($j=5$) and boundary ($j=0$) sites respectively. In either case, following the initial coherent evolution, we see that the dissipative site rapidly relaxes to the spin-$\downarrow$ state due to the local dissipation enacted by the jump operator $L_j = S_j^-$. On the other hand, we observe that localization counteracts the spread of the incoherent decay process by the long-range XY interaction, leading to slow relaxation of the local magnetization away from the dissipative site. Moreover, we find that the rate of decay decreases as a function of distance from the dissipative site in either of the cases considered here. In the limit of strong disorder and weak dissipation, this behavior stems from the fact that the local decay rate at site $k$ is approximately set by the wave-function overlap between the eigenstate localized around the dissipative site $j$ and the state localized at site $k$, which decreases with $|j-k|$ if the system is in a localized phase (see discussion below).

\begin{figure}[t]
\centering
\includegraphics[clip, trim=.5cm 7cm 0.5cm 8cm,width= .45\textwidth]{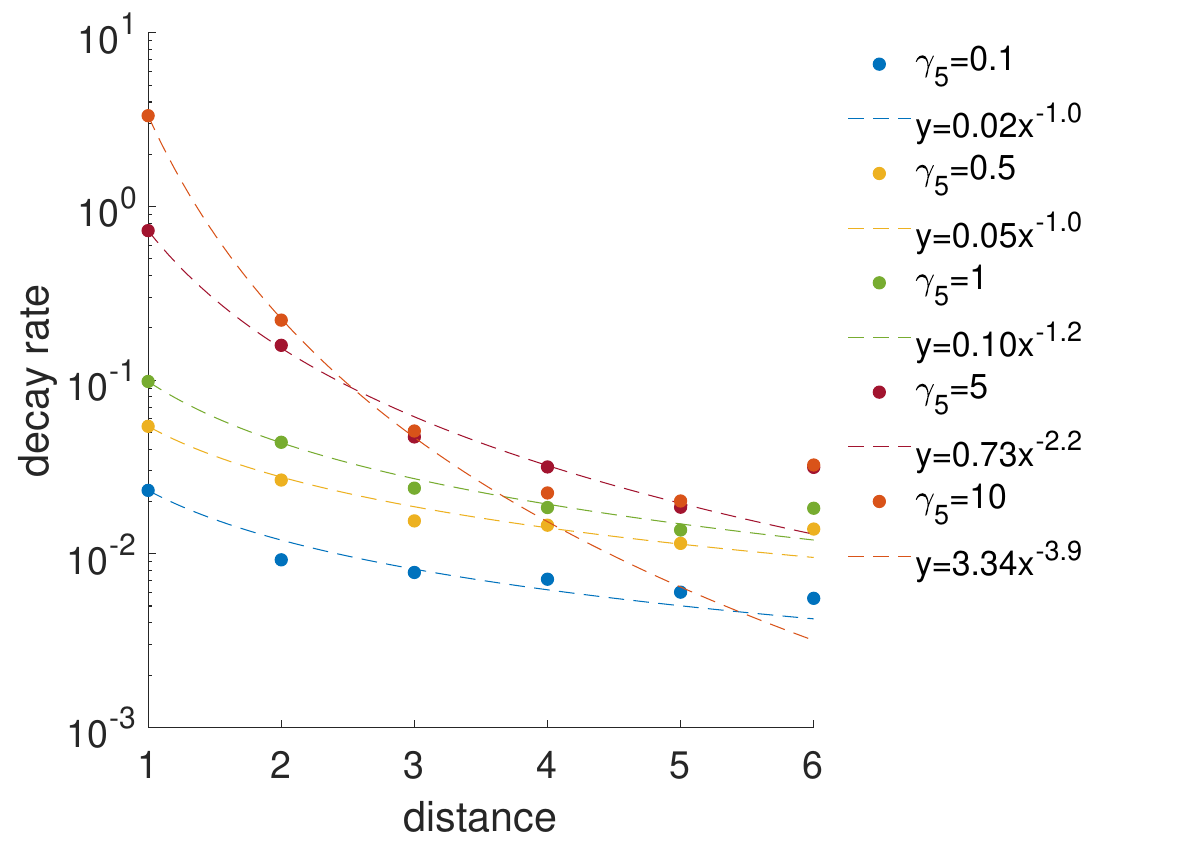}
\includegraphics[clip, trim=.5cm 7cm 0.5cm 8cm,width= .45\textwidth]{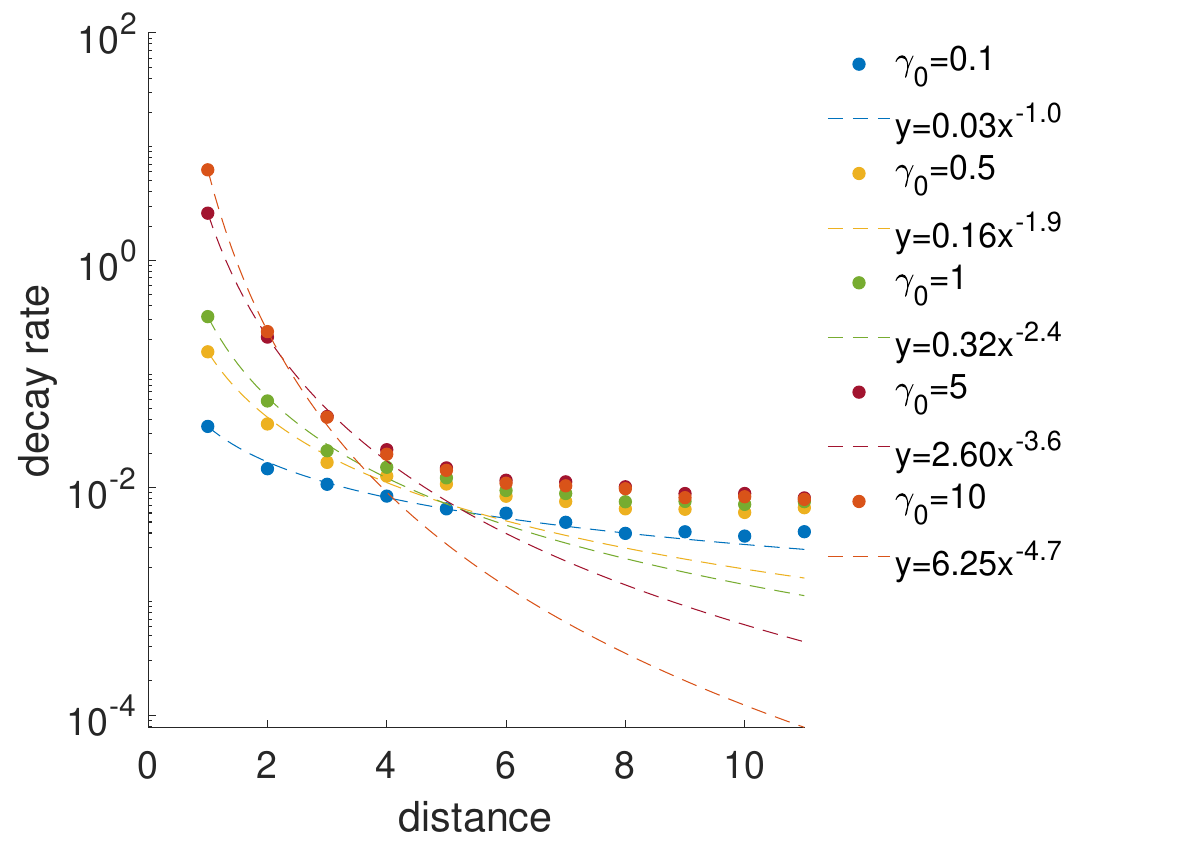}
\caption{\small Local exponential decay rates extracted from the dynamics of $\langle S_k^z \rangle$ in an $L=11$ site system with unit coupling ($\bar{J}=1$) and moderate disorder ($W=10$) as a function of $|j-k|$, the distance from the lossy site for various dissipation strengths. The top (bottom) panel shows results for local dissipation at the central $j=5$ (edge $j=0$) site with strength $\gamma_5$ ($\gamma_0$). The dashed curves indicate fits to power law behaviour using the least squares method.}
  \label{fig:decays}
\end{figure}

We observe qualitatively similar behavior for local dissipation rates $0.1  \leq \gamma_j \leq 10$, in that the relaxation rate decreases with increasing distance from the dissipative site. However, the precise functional form of the local decay rate as a function of distance from the dissipative site quantitatively depends on the strength of the dissipation. Specifically, in Fig.~\ref{fig:decays} we show the relationship between the fitted exponential decay rates for $\langle S_i^z \rangle$ and the distance $|j-k|$ from the lossy site for various dissipation rates $\gamma$. For weak dissipation $\gamma \sim 0.1$, we find that the local decay rate falls with the distance from the lossy site according to a power law $\sim 1/|j-k|$. However, this power-law relationship fails when the dissipation strength exceeds a certain value ($\gamma=1$ at the central site and $\gamma=0.1$ at the edge site). This is consistent with the expectation that, in the regime of strong dissipation, such a system can enter a quantum Zeno localized regime where the effective dissipation rate for the non-dissipative sites is significantly suppressed~\cite{popkov2018effective}. 

\subsection{Semi-analytic argument: Global relaxation via local dissipation}

Here, we provide a simple argument for why a localized system is expected to relax at late-times in the presence of local dissipation at a single site (see Ref.~\cite{lezama2021}, which provides a similar argument for local dephasing). Consider a 1D free-fermion chain with long-range hopping and on-site disorder, with the Hamiltonian
\beq
\label{eqn:nonham}
H = \sum_{i\neq j}\frac{J}{|i-j|^\alpha}c_i^\dagger c_j + \sum_j \epsilon_j c_j^\dagger c_j \, ,
\eeq
where $c/c^\dagger$ are local creation/annihilation operators that satisfy canonical anti-commutation relations, $\epsilon_i \in [-\epsilon,\epsilon]$ and we take $\alpha > 2$ in order for the system to localize for non-zero disorder~\cite{anderson1958,levitov1990,aleiner2011}. While there is numerical evidence suggesting the existence of a many-body localized phase above a critical disorder strength once interactions are added to this system (for $\alpha > 2$)~\cite{yao2014many,wu2016,mirlin2018}, since we will argue that even the Anderson insulator is unstable to the addition of single-site dissipation, this should clearly hold for the many-body localized phase as well (the destablization of MBL due to the inclusion of a single thermal grain in an otherwise localized system is referred to as the avalanche instability~\cite{de2017stability,luitz2017,thiery2018avalanche}).

The Lindblad equation that includes local dissipation at a single site $k$ at a rate $\gamma$ is given by: 
\beq
\partial_t \rho = -i [H,\rho] + \gamma \left( c_j \rho c_j^\dagger - \frac{1}{2} \{c_j^\dagger c_j \} \right) \, ,
\eeq
where the jump operator $L_j = c_j$ induces incoherent particle-loss at site $j$. Note that the spin-$1/2$ XY model given by Eq.~\eqref{eqn:ham} maps precisely to this non-interacting free-fermion model with particle-loss via a Jordan-Wigner transformation if the dissipation $S_j^-$ acts on the first $j=0$ or last $j=L-1$ site, but is generically interacting. From the above master equation, we can derive the time-evolution for expectation values of local observables: 
\beq
\partial_t \langle \mathcal{O} \rangle = Tr[\mathcal{O}\partial_t \rho] = i \langle [H,\mathcal{O}] \rangle - \gamma Re\langle c_j^\dagger [c_j,\mathcal{O}] \rangle \, .
\eeq
We can express the single-particle eigenstates of the non-interacting Hamiltonian Eq.~\eqref{eqn:nonham} as
\beq
b_\ell = \sum_j \alpha_{\ell,j} c_j \, , \quad c_j = \sum_{\ell} \alpha_{\ell,j}^* b_\ell \, ,
\eeq
and since the mode occupation numbers $n_\ell = b_\ell^\dagger b_\ell$ commute with the Hamiltonian (by definition), we obtain the following rate equation that includes the effect of the local dissipation at site $j$:
\beq
\partial_t \langle n_\ell \rangle = -\gamma Re\langle c_j^\dagger [c_j,n_\ell] \rangle = -\gamma Re[\alpha_{\ell,j}^* \sum_k \alpha_{k,j} \langle b_k^\dagger b_\ell \rangle] \, .
\eeq
For a localized system, we can assume that the modes are uncorrelated: 
\beq
\langle b_k^\dagger b_\ell \rangle \approx \delta_{k,\ell} \langle n_\ell \rangle \, ,
\eeq
which gives the simplified rate equation
\beq
\partial_t \langle n_\ell \rangle = - \gamma |\alpha_{\ell,j}|^2 \langle n_\ell \rangle \,.
\eeq
This clearly demonstrates that even an Anderson localized system will exhibit exponential decay of the local population at sites $\ell$ that are distinct from the lossy site. However, the rate of decay scales is set by the coefficients $\alpha_{\ell,j}$ which describe the wavefunction overlap between the eigenstates localized at site $\ell$ and the eigenstate localized at the dissipative site $j$. Since this overlap is determined by the localization length, the decay rate will exponentially decrease as a function of the distance $|j-\ell|$, leading to slow relaxation dynamics for sites which are well-separated from the lossy site. While this can be shown analytically for the non-interacting (Anderson) case, we have also verified numerically that this exponential decrease away from the dissipative site is present even once short-range interactions are turned on (in the limit of strong disorder and weak dissipation (see also Ref.~\cite{toniolo2024}).

From our numerical analysis for the long-range interacting XY chain, in the limit of strong disorder and weak dissipation, we observe that the overlap $|\alpha_{\ell,j}|^2 \sim 1/r$, which suggests that in this limit only a vanishing density of sites near the dissipative site will decay while the others will continue to evolve coherently. At the opposite extreme, our numerics suggest that strong dissipation can drive an effective quantum Zeno dynamics, such that at time-scales long compared to $\gamma$, the non-dissipative sites exhibit decay rates that are significantly smaller than $\gamma$~\cite{popkov2018effective}. A detailed investigation of the full many-body dynamics of this disordered, long-range interacting system subject to local dissipation will be the subject of a forthcoming work.

\begin{table*}[t]
\caption{Conversion between model parameters $\hbar=1$, $\bar{J}=1$, $r=1$, $W=10$, and $\gamma=0.1$, to physical dimensions $\bar{J}$=75 GHz $\mu m^3$ at different particle densities.}
\label{Table:units}
\begin{tabular}{ccccccc}
\toprule
$\rho$ & nearest-neighbor distance  & \hspace{20 pt} unit time \hspace{20 pt}  & \hspace{20 pt}  W   \hspace{20 pt} &        \hspace{20 pt}       $\gamma$ \hspace{20 pt}  & \hspace{5 pt} decay time \hspace{5 pt}  & \hspace{5 pt} 500 unit time \\ 
$(\mu m^{-3})$ &  ($\mu$m) & 1/(2$\pi$*75GHz)  (ns) &  (GHz) &               (GHz) & (ns) & (ns) \\ 
\midrule
1                 & 1.0                                & 0.002                           & 750     & 7.5                         & 0.13            & 1.1                \\
0.5               & 1.3                                & 0.004                           & 375     & 3.75                        & 0.27            & 2.1                \\
0.1               & 2.2                                & 0.021                           & 75      & 0.75                        & 1.33            & 10.6               \\
0.01              & 4.6                                & 0.212                           & 7.5     & 0.075                       & 13.33           & 106.2 \\
\bottomrule
\end{tabular}
\end{table*}

\subsection{Dimensional analysis}  

Here, we provide a dictionary between the parameters used in the toy model and the experimental parameters (in physical units). For the numerical simulations, the dimensions scale with the density of spins. Table \ref{Table:units} provides the conversion from units in which $\hbar=1$, $\bar{J}=1$, $r=1$, $W=10$, and $\gamma=0.1$, to the real physical units with $\bar{J}$=75 GHz $\mu m^3$ and for different densities of spins; here, a density of 0.5 $\mu$m$^{-3}$ corresponds to an average nearest neighbour distance of 1.26 $\mu$m. At this density, disorder of $W=10$ spans a width of 375 GHz, $J_{ij}=\bar{J}/r_{ij}^3 =0.5$ refers to a coupling width of 38 GHz, and $\gamma_j =0.1$ describes a decay time of 270 ps. Under these conditions, the full scale of $t*\bar{J}=500$ defines a total elapsed time of about 2 ns. 

Thus, we have shown that this simple toy model qualitatively captures the relaxation dyanmics observed in the experiment. Namely, the introduction of a moderate rate of decay at a single site within a dipolar system, that is otherwise localized by strong diagonal disorder, can drive global dissipation across the entire system, where each dipole exhibits a rate of decay that varies algebraically with its distance from the locally dissipative site. 

\section{Conclusions}  

The molecular ultracold plasma that forms following the avalanche and quench of a state-selected Rydberg gas of nitric oxide undergoes an arrested relaxation that exhibits the signature of an enduring prethermal phase.  These dynamics begin with double-resonant excitation, which prepares an initial ensemble composed of molecules in a single selected state of principal quantum number, $n_0$.  Penning ionization, followed by an avalanche of electron-Rydberg collisions, forms a plasma of NO$^+$ ions and weakly bound electrons, in which electron collisions drive a residual population of $n_0$ Rydberg molecules to a state of very high-$\ell$.  Predissociation further depletes this plasma of low-$\ell$ molecules, purifying the high-$\ell$ ensemble.  For a wide range of initial conditions, this system enters a critical phase in which a density of NO$^+$ and electrons balances a population of Rydberg molecules. 

This system ceases to progress toward a statistically equilibrated state composed largely of ${\rm N~(^4S)}$ and  ${\rm O~(^3P)}$ atoms.  Such thermalization requires a flow of energy bound in the Coloumb separation of electrons and NO$^+$ ions to the electronic degrees of freedom of the neutral molecule.  This cannot occur without Rydberg electron penetration.  In this isolated system, an extraordinary angular momentum gap between the plasma states of $n \approx \ell$, with measured $n>200$, and penetrating states of $\ell = 0, ~1$ and 2, apparently blocks this relaxation for longer than a millisecond. It remains an exciting open question to better understand the relation, if any, between prethermal MBL and the localized prethermal state observed here. Drawing a precise connection between the two would require a thorough investigation of the relaxation timescale as a function of the disorder strength, which is expected to be exponential for prethermal MBL.

Evolving through the critical phase, the electrons that balance the NO$^+$ charge behave as though localized in the prethermal phase and do not play an effective role in bridging this gap.  However, the application of an RF field with an amplitude as small as 200 mV cm$^-1$ promotes a dramatic degree of relaxation.  The density dependence of this effect clearly associates it with RF-initiated electron-collisional $\ell$-mixing.  

On an entirely different scale, promoting a quantum-state transition in an exceedingly small fraction of the molecules in the prethermalized ensemble acts with an even greater effect to drive the entire system toward equilibrium.  We ascribe this to an introduction of dissipative character to a small fraction of the states in the prethermally localized ensemble.  In time, this character spreads to bridge the angular momentum gap and cause the entire ensemble to thermalize.  Finally, we qualitatively reproduce similar dynamics within the Lindblad formalism for a toy model of an open quantum spin chain that includes on-site disorder, long-range interactions, and local dissipative coupling to a Markovian reservoir. The preliminary numerical results we have presented here suggest that this simple 1D model provides a rich playground for investigating the nontrivial interplay between local decoherence and constrained quantum dynamics in a prethermalized Rydberg system.

\begin{acknowledgements}

This work was supported by the US Air Force Office of Scientific Research (AFOSR Grant No. FA9550-17-1-0343), together with the Natural Sciences and Engineering Research Council of Canada (NSERC Grant No. RGPIN- 2019-04242), the Canada Foundation for Innovation (CFI) and the British Columbia Knowledge Development Fund (BCKDF). This material is based upon work supported by the Sivian Fund at the Institute for Advanced Study and the U.S. Department of Energy, Office of Science, Office of High Energy Physics under Award Number DE-SC0009988 (A.P.).

\end{acknowledgements}

\bibliography{references,AFOSR_refs,AF_bib,prethermal,AP_ref}

\end{document}